\documentclass[12pt]{article}
\usepackage{tabularx} 
\usepackage{amsmath}  
\usepackage{amsfonts}
\usepackage{authblk}
\usepackage{stmaryrd}
\usepackage{parskip}
\usepackage{floatrow}
\usepackage{todonotes}
\usepackage{subcaption}
\usepackage{graphicx}
\usepackage{amsmath}
\usepackage{mdframed}
\usepackage{mathtools}
\usepackage{amsfonts}
\usepackage{amssymb}
\usepackage[section]{placeins}
\usepackage{placeins}
\usepackage{graphicx} 
\usepackage[margin=1.0in]{geometry} 
\usepackage[ruled,vlined]{algorithm2e}
\usepackage{cite} 
\usepackage[final]{hyperref} 
\hypersetup{
	colorlinks=true,       
	linkcolor=blue,        
	citecolor=blue,        
	filecolor=magenta,     
	urlcolor=black,   
}

\makeatletter
\AtBeginDocument{%
  \expandafter\renewcommand\expandafter\subsection\expandafter{%
    \expandafter\@fb@secFB\subsection
  }%
}
\makeatother

\title{A Multi-Resolution Approach \\ to Hydraulic Fracture Simulation}

\author[a]{Andre Costa}
\author[b]{Matteo Cusini}
\author[c]{Tao Jin}
\author[b]{Randolph Settgast}
\author[a, *]{John Dolbow}

\affil[a]{Department of Mechanical Engineering and Material Science, Duke University, Durham, 27708, NC, USA} 

\affil[b]{Atmospheric, Earth and Energy Division, Lawrence Livermore National Laboratories, Livermore, 94550, CA, USA} 

\affil[c]{Department of Mechanical Engineering, University of Ottawa, Ottawa, K1N6N5, ON, Canada} 

\affil[*]{Corresponding author: John Dolbow, john.dolbow@duke.edu}
  
\begin{document}

\maketitle

\begin{abstract}
We present a multi-resolution approach for constructing model-based simulations of hydraulic fracturing, wherein flow through porous media is coupled with fluid-driven fracture.  The approach consists of a hybrid scheme that couples a discrete crack representation in a global domain to a phase-field representation in a local subdomain near the crack tip. 
The multi-resolution approach addresses issues such as the computational expense of accurate hydraulic fracture simulations and the difficulties associated with reconstructing crack apertures from diffuse fracture representations. 
In the global domain, a coupled system of equations for displacements and pressures is considered. The crack geometry is assumed to be fixed and the displacement field is enriched with discontinuous functions. Around the crack tips in the local subdomains, phase-field sub-problems are instantiated on the fly to propagate fractures in arbitrary, mesh independent directions. The governing equations and fields in the global and local domains are approximated using a combination of finite-volume and finite element discretizations.  The efficacy of the method is illustrated through various benchmark problems in hydraulic fracturing, as well as a new study of fluid-driven crack growth around a stiff inclusion.  

\end{abstract}

\section{Introduction}

A wide range of approaches for model-based simulations of hydraulic fracturing have been developed over the past several decades \cite{adachi2007computer, lecampion2018numerical}.  These range from production-level reservoir modeling tools such as  ResFrac\cite{mcclure2017three, mcclure2018resfrac}, to GEOS\cite{settgast2012simulation, settgast2014simulation, settgast2017fully}, PyFrac\cite{zia2020pyfrac}, and others.  Many of the models and associated codes assume fracture networks that remain planar, but in recent years strides have been made towards modeling cracks that evolve in arbitrary ways in response to fluid-driven loads. High-resolution models for complex fracture evolution generally fall into two categories: sharp interface models that explicitly model the fracture surface, and diffuse crack approaches that effectively smear the geometry over the underlying grid or mesh.  Techniques that represent the crack as a sharp interface can be advantageous when the fracture configuration is relatively simple, but representing complex geometric evolution can be challenging  \cite{gupta2014simulation, gupta2018coupled, shauer2022three}.  By contrast, diffuse crack models offer more flexibility for representing complex fracture evolution, but introduce other challenges such as the lack of a well-defined fracture surface and increased computational expense\cite{heider2021review}.  In this work, we introduce a multi-resolution scheme for hydraulic fracturing simulation that makes use of both sharp and diffuse crack representations within a single framework. The objective is to establish a methodology that makes use of the advantageous aspects of both sharp and diffuse crack models while circumventing some of the drawbacks.  

Over the past several decades, the phase-field model for fracture \cite{francfort1998revisiting, bourdin2000numerical, karma2001phase} has emerged as a promising approach for constructing robust simulations of complex crack evolution.  The method has shown considerable success for simulating fracture evolution in quasi-brittle materials, and there have been several recent efforts to extend the approach to hydraulic fracturing. In what follows, we review some prior works of particular relevance to the current manuscript.  For additional references in this topic, we refer the reader to the recent review by Heider \cite{heider2021review}.  

The first attempts towards a phase-field model for hydraulic fracture began with extensions of the traditional phase-field model to pressurized cracks, as in Bourdin et al. \cite{bourdin2012variational} and Wheeler et al. \cite{wheeler2014augmented}. Subsequently, fluid flow in the fractures, and also poromechanics were considered. Miehe et al. \cite{miehe2015minimization, miehe2016phase} developed a thermodynamically consistent framework, from minimization principles, to couple poromechanics, fluid-flow and phase-field fracture. The flow problem was modeled via the Darcy's equation, containing a permeability coefficient that used the phase-field variable and the crack opening to mimic the cubic relationship from the lubrication theory in the crack region. Mikelic et al. \cite{mikelic2015phase1, mikelic2015phase2} developed a model that separated the domain into fracture and reservoir, by using the phase-field variable as an indicator function. They also considered the flow inside the fracture as a Darcy flow, but their model treated the fracture as a three-dimensional entity, which led to a different permeability tensor compared to \cite{miehe2015minimization, miehe2016phase}.  Yet another approach concerns the work of Wilson and Landis \cite{wilson2016phase}, who proposed a model that included fluid velocities as primary variables. This allowed for a more detailed description of the flow within the fracture, which was modeled by a Brinkman-type equation \cite{brinkman1949calculation}. The phase-field parameter acted as an indicator of the flow regime, between Darcy flow (away from cracks) and Stokes flow (inside cracks).  Finally, the recent work of Chukwudozie et al.\ \cite{chukwudozie2019variational} presented a different model, wherein the lubrication theory equations were included in the weak form by means of a $\Gamma$-convergent regularization. 

The use of a phase-field to represent a fracture network in a diffuse manner certainly facilitates the representation of complex geometric evolution, including crack branching and merging. However, it also requires the use of meshes or grids that are capable of resolving the regularization length, making these approaches computationally expensive. In the specific case of hydraulic fracturing, another challenge concerns the crack opening or aperture, a field that is tightly coupled with the fluid pressure within fractures. In a phase-field setting, due to the lack of an explicit crack surface, extracting the aperture or accounting for its effects requires additional considerations.  All of the aforementioned  works present some way to account for the aperture within a diffuse setting, but the robustness of these approaches remains unclear \cite{lecampion2018numerical}. For a review of the most frequently used methods to calculate the crack aperture from phase-field simulations, see the recent work of Yoshioka et al.\ \cite{yoshioka2020crack}.

Outside of the context of hydraulic fracturing, some researchers in the phase-field community have developed ``hybrid" approaches, wherein the phase-field formulation was combined with a sharp crack representation. The motivation for these approaches varies, from ``cutting" the mesh to remove artificial traction transmission and circumvent element distortion \cite{geelen2018optimization} to reducing the overall computational cost \cite{giovanardi2017hybrid, muixi2021combined}. In the work of Giovanardi et al.\ \cite{giovanardi2017hybrid}, phase-field subproblems in the vicinity of  crack tips were used to propagate a global, discrete crack. The eXtended Finite Element Method (XFEM)\cite{moes1999finite} was used to place fracture discontinuities in the displacement field within the background global mesh. More recently, Muixi et al.\ \cite{muixi2021combined} created an approach that uses the phase-field method only at the crack tips, and XFEM in the rest of the domain. In contrast to \cite{giovanardi2017hybrid}, there is no overlap of the representations in crack tip areas.

The success of these hybrid approaches for purely mechanical cases opens the door for their extension to hydraulic fracturing.  Such approaches are appealing because in principle they can circumvent the need for a complicated reconstruction of the crack opening from the phase-field.  This area is relatively unexplored, although there have been some recent efforts that are similar in spirit, such as the recent work of Sun et al.\ \cite{sun2020hybrid}.  They developed a Finite Element-Meshfree method to represent the crack surfaces in a discrete fashion. The computed displacement field was used to obtain a driving force which was employed within a phase-field evolution equation near the crack tips. This approach eliminated the need for the reconstruction of crack openings from the diffuse crack representation, but it also largely decoupled the phase field from the equations governing the force balance near the crack tips.  

The approach presented in this work, which we refer to as a multi-resolution method, extends the concept of a hybrid phase-field method to hydraulic fracturing. It discretizes the problem at a global level using a sharp interface approach based on the Embedded Finite Element Method of  Cusini et al.\ \cite{cusini2021simulation}.  It then approaches the simulation of fracture evolution by coupling the global fields with a phase-field fracture problem posed over a subdomain in the vicinity of the crack front.  This approach has several advantages.  The crack aperture and flow inside the fracture are handled at the global scale using techniques that work well and are efficient when the crack geometry is known.  Then the phase-field model is employed in the subdomain  to effectively update the crack geometry.  This framework lends itself to incorporation with a wide range of existing hydraulic fracturing solvers, as the phase-field sub-problem is agnostic about the type of numerical treatment used in the global domain. 

The paper is structured as follows. In Section \ref{formulation}, we present the governing equations and constitutive assumptions for both hydraulic fracturing and phase-field for fracture. In Section \ref{numerics}, we propose our multi-resolution framework and present numerical schemes to discretize both the hydrofracture and the phase-field subproblems. In Section \ref{results_section}, we apply our method to study hydraulic fracturing in some simple scenarios, in order to verify its accuracy. These include the well-known KGD\cite{geertsma1969rapid, zheltov19553} problems and a case of non-planar fracture propagation around a stiff inclusion. Finally, in Section 5, we provide a summary and some concluding remarks.

\section{Model formulation}\label{formulation}

In this section, we describe the models that are used to develop our multi-resolution scheme.  We start by presenting the governing equations to model hydraulic fracture in poroelastic rock that forms the basis for the solver at the global scale.  We then describe a phase-field model for fracture that forms the basis for the solver used in local subdomains near the crack tips.  

\subsection{Governing equations for hydraulic fracture}

\begin{figure}[h]
    \centering
    \includegraphics[width=15cm]{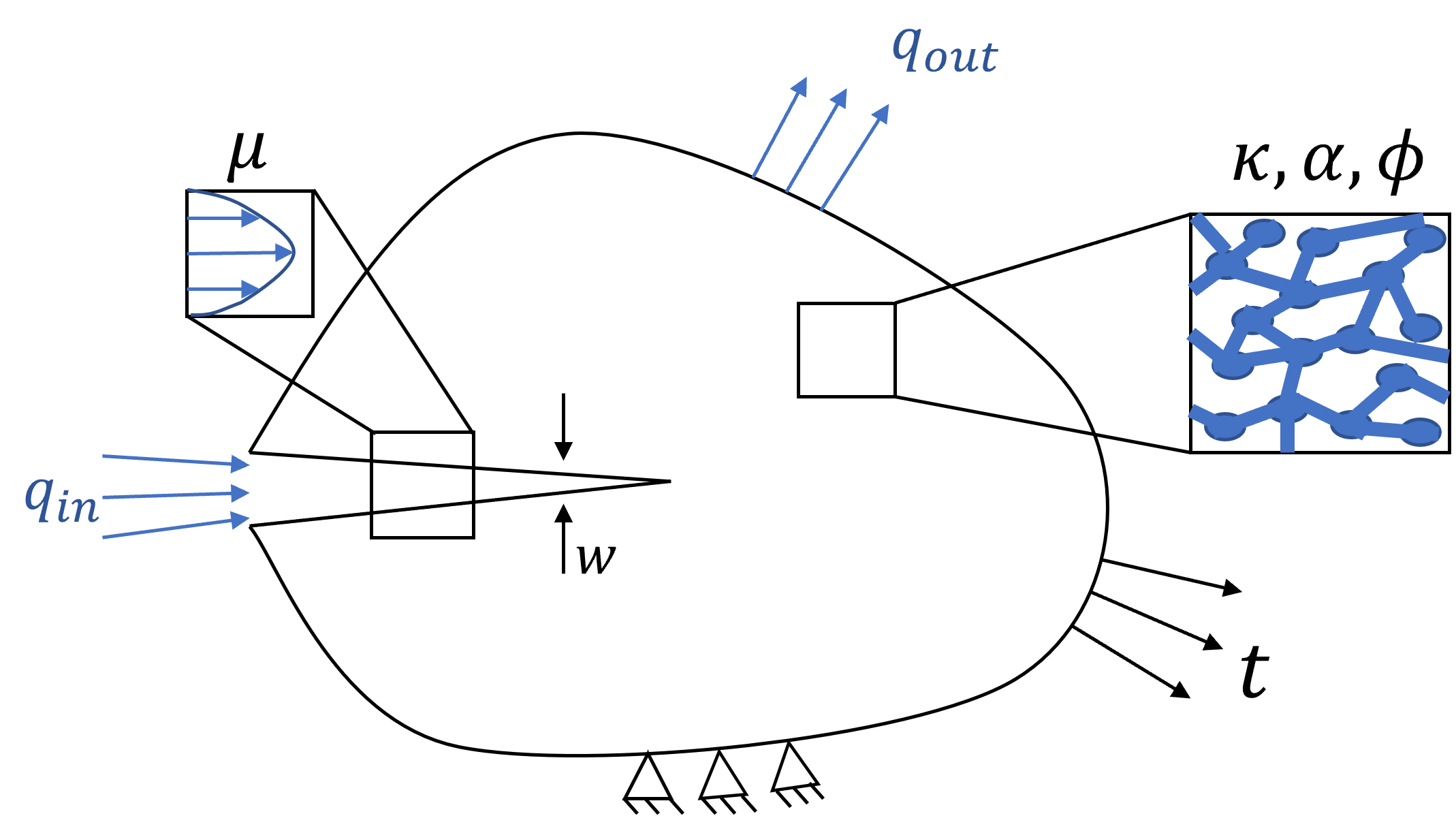}
    \caption{Schematic of a poroelastic rock with a fracture, inspired by \cite{landis2016birs}.}
    \label{fig:potato_rock}
\end{figure}

We consider a model that couples flow and elastic deformation in a porous media with evolving fracture surfaces.  Consider the domain $\Omega$ consisting of a porous rock, that is fully saturated with a single-phase Newtonian fluid. The external boundary is composed of both traction $\partial \Omega_t$ and displacement $\partial \Omega_u$ surfaces, viz.\  $\partial \Omega = \partial \Omega_t \cup \partial \Omega_u$. 

The domain contains fractures $\Gamma$, as shown in Figure \ref{fig:potato_rock}.  For simplicity, we assume quasi-static conditions and small-strain kinematics. Neglecting inertial effects, the balance of linear momentum reads
\begin{equation}\label{linear momentum balance}
    \nabla \cdot \boldsymbol\sigma + \textbf{b} = \boldsymbol 0\ \text{on } \Omega\setminus\Gamma,
\end{equation}
where $\boldsymbol\sigma$ denotes the total Cauchy stress and $\textbf{b}$ the body force. The mechanical boundary conditions are given by
\begin{equation}\label{traction bc}
    \boldsymbol \sigma \cdot \textbf{n} = \textbf{t} \text{ on } \partial \Omega_t,
\end{equation}
\begin{equation}\label{displacement bc}
    \textbf{u} = \overline{\textbf{u}} \text{ on } \partial \Omega_u,
\end{equation}

where $\textbf{n}$ denotes the normal direction at any point on $\partial\Omega$,  $\textbf{t}$ are the applied tractions, and $\overline{\textbf{u}}$ denotes the prescribed displacements.

Fluid flow within the cracks gives rise to pressure loads on the crack surfaces, translating into the boundary condition 
\begin{equation}\label{fracture boundary condition}
    \boldsymbol\sigma^+\cdot \textbf{n}_{\Gamma} = -\boldsymbol\sigma^-\cdot \textbf{n}_{\Gamma} = -p_f\textbf{n}_{\Gamma} \text{ on } \Gamma,
\end{equation}
where $p_f$ is the pressure inside the crack and $\textbf{n}_{\Gamma}$ is the normal vector to $\Gamma$ at any point.  In this work, we assume that fractures are open and neglect contact conditions on crack faces.  For additional considerations to account for closed fractures, see the model described in Cusini et al.\cite{cusini2021simulation}.

In terms of the fluid flow, the fluid velocity $\textbf{v}_m$ in the matrix is governed by the mass balance 
\begin{equation}\label{mass balance matrix}
    \dfrac{\partial(\rho \phi)}{\partial t} + \nabla \cdot (\rho \textbf{v}_m) = Q_m + Q_{mf}\ \text{on } \Omega\setminus\Gamma.
\end{equation}
where $\rho$ denotes the fluid density and $\phi$ is the porosity, and $Q_m$ is a prescribed source term in the matrix. The source term $Q_{mf}$ accounts for the exchange of fluid between the matrix and the fractures.
The boundary is partitioned into pressure $\partial \Omega_p$ and flux $\partial\Omega_q$ portions, such that $\partial \Omega = \partial \Omega_p \cup \partial \Omega_q$, and the following boundary conditions are applied,

\begin{equation}\label{flux bc}
    \rho \textbf{v}_m \cdot \textbf{n} = \textbf{q} \text{ on } \partial \Omega_q,
\end{equation}

\begin{equation}\label{pressure bc}
    p_m = \overline{p}_m \text{ on } \partial \Omega_p,
\end{equation}

where $p_m$ denotes the pore-pressure and  $\overline{p}_m$ is a prescribed pressure.

In a comparable manner, the fluid velocity $\textbf{v}_f$ within fractures is governed by the mass balance
\begin{equation}\label{mass balance fracture}
    \dfrac{\partial(\rho w)}{\partial t} + \nabla_{\Gamma} \cdot (\rho w \textbf{v}_f) = Q_f + Q_{fm}\ \text{on } \Gamma,
\end{equation}
where $w$ denotes the normal fracture aperture, $Q_f$ is a prescribed source term within the fracture, and $Q_{fm}$ accounts for the exchange of fluid between the fracture and the matrix\footnote{the flux interactions $Q_{fm}$ and $Q_{mf}$ are modeled as in classical well models, following \cite{hajibeygi2011hierarchical}. This ensures the balance of mass between the fractures and matrix $\int_V Q_{mf} dV = -\int_{\Gamma}Q_{fm}d\Gamma$.}. In the above, $\nabla_{\Gamma}$ indicates a gradient operator taken on the lower dimensional manifold  $\Gamma$. Boundary conditions similar to \eqref{flux bc} and \eqref{pressure bc} can also be applied.

Constitutive relationships are required to close the system and tie the stresses to the displacements $\textbf{u}$ and the fluid velocities to the pressures. In particular, we adopt the basic assumptions of Biot's theory of poroelasticity \cite{biot1941general}, Darcy's law for the flow in the matrix, and a lubrication theory approximation for the flow in fractures. This gives rise to the following set of constitutive relationships for the stress, porosity, and velocities:

\begin{equation}\label{biot law}
    \boldsymbol \sigma = \mathbb{C}:\boldsymbol\epsilon(\textbf{u}) - \alpha p_m \mathbb{I} \ \text{on } \Omega\setminus\Gamma,
\end{equation}

\begin{equation}\label{linear porosity}
    \dot{\phi} = \alpha \nabla \cdot \dot{\textbf{u}} + \dfrac{\dot{p_m}}{N} \ \text{on } \Omega\setminus\Gamma,
\end{equation}

\begin{equation}\label{darcy law}
    \textbf{v}_m = -\dfrac{\kappa}{\mu}\nabla p_m \ \text{on } \Omega\setminus\Gamma,
\end{equation}

\begin{equation}\label{cubic law}
    \textbf{v}_f = -\dfrac{w^2}{12\mu}\nabla_{\Gamma} p_f \ \text{on } \Gamma.
\end{equation}

In the above, $p_m$ is the pore-pressure, $\mathbb{C}$ is the fourth-order isotropic tensor of drained elastic moduli,  $\boldsymbol\epsilon(\textbf{u})$ is the mechanical strain, $\alpha$ is the Biot coefficient, and $\mathbb{I}$ is the second-order identity tensor. The temporal evolution of the porosity is governed by the rate of dilatation and time rate of change in the pore pressure, as modulated by the modulus $N$.  The fluid velocity in the matrix is related to the gradient of the pressure through the ratio of the intrinsic permeability $\kappa$ to the viscosity $\mu$.    

The fluid is assumed to be linearly compressible.  For both the fluid in the fractures and the matrix, this implies that the density is updated from its reference value $\rho_{ref}$ based on the change in pressure according to
\begin{equation}\label{poorly compressibility}
    \rho = \rho_{ref} \left(1 + \dfrac{p  - p_{ref}}{K_F}\right) \ \text{on } \Omega,
\end{equation}
where $p_{ref}$ denotes a reference value for the pressure, and $K_F$ is the fluid bulk modulus. 

Finally, the initial conditions for the displacements and pressures are given by

\begin{equation}\label{u_ic}
    \textbf{u}(\textbf{x},0) = \textbf{u}^0  \ \text{on } \Omega\setminus\Gamma,
\end{equation}

\begin{equation}\label{pm_ic}
    p_m(\textbf{x},0)= p^0_m  \ \text{on } \Omega\setminus\Gamma,
\end{equation}

\begin{equation}\label{pf_ic}
    p_f(\textbf{x},0) = p^0_f  \ \text{on } \Gamma.
\end{equation}

For a given crack geometry, the combination of equations \eqref{linear momentum balance}, \eqref{mass balance matrix} and \eqref{mass balance fracture}, with constitutive assumptions \eqref{biot law} - \eqref{poorly compressibility}, boundary conditions \eqref{traction bc}-\eqref{fracture boundary condition},\eqref{flux bc},\eqref{pressure bc} and initial conditions \eqref{u_ic}-\eqref{pf_ic} leads to a system of equations whose solution can be approximated by many different numerical methods.
What remains is a model to describe the evolution of the crack geometry.  In the next subsection, we present the governing equations for a phase-field method for fracture, a regularized approach for representing fractures and their evolution.

\subsection{The phase-field method for fracture}

The phase-field method for fracture started as an approximation \cite{bourdin2000numerical} to the variational approach for fracture by Francfort and Marigo \cite{francfort1998revisiting} for the quasi-static propagation of fracture in brittle materials. For the method adopted in this work, we follow the work of Chukwudozie et al.\ \cite{chukwudozie2019variational} and associate the following total energy to a crack configuration $\Gamma$ in a poroelastic brittle solid $\Omega$:
\begin{multline}\label{variational approach to fracture}
    \mathcal{E}(\textbf{u}, p_m, p_f, \Gamma) = \int\limits_{\Omega \setminus \Gamma}W(\boldsymbol{\epsilon}(\textbf{u}), p_m)d\Omega - \int\limits_{\partial\Omega_N} \textbf{t} \cdot \textbf{u} ds
    - \int\limits_{\Omega\setminus \Gamma} \textbf{b} \cdot \textbf{u} d\Omega \\
    + \int\limits_{\Gamma}p_f \llbracket \textbf{u} \cdot \textbf{n}_{\Gamma} \rrbracket ds + G_c \mathcal{H}^{n-1}(\Gamma),
\end{multline}
where the tractions applied to the boundary are denoted by $\textbf{t}$, the normal to the crack is denoted by $\textbf{n}_{\Gamma}$ and $\mathcal{H}^{n-1}(\Gamma)$ is the $n-1$ dimensional Hausdorff measure of $\Gamma$. 

The strain energy density $W(\boldsymbol{\epsilon}(\textbf{u}), p_m)$ is postulated as,

\begin{equation}
    W(\boldsymbol{\epsilon}(\textbf{u}), p_m) = \dfrac{1}{2}\left( \boldsymbol\epsilon(\textbf{u}) - \dfrac{\alpha}{nK} p_m\mathbb{I}\right) : \mathbb{C} : \left( \boldsymbol\epsilon(\textbf{u}) - \dfrac{\alpha}{nK} p_m\mathbb{I}\right),
\end{equation}

with $K$ denoting the bulk modulus and $n$ the system's dimension (2 or 3). The phase-field regularization, based on the Ambrosio-Tortorelli \cite{ambrosio1990approximation} functional is then performed by the introduction of the damage parameter $d$ and the regularization length $\ell$,

\begin{multline}\label{Poroelastic PF funcional}
    \mathcal{E}_{\ell}(\textbf{u},d,p_m,p_f) = \int\limits_{\Omega }\widetilde{W}(\boldsymbol{\epsilon}(\textbf{u}), p_m, d)d\Omega - \int\limits_{\partial\Omega_N}\textbf{t} \cdot \textbf{u} ds
    - \int\limits_{\Omega} \textbf{b} \cdot \textbf{u} d\Omega
    + \int\limits_{\Omega}p_f \textbf{u} \cdot \nabla d d\Omega \\
    + \dfrac{G_c}{c_0}\int\limits_{\Omega }\left(\dfrac{\zeta(d)}{\ell} + \ell\nabla d\cdot\nabla d  \right)d\Omega,
\end{multline}

where the function $\zeta(d)$ is the local dissipation function, which is usually taken as $\zeta(d) = d$ or $d^2$. The constant $c_0$ is given by $c_0 = 4\int_0^1 \sqrt{\zeta(z)}dz$ and the regularized strain energy is defined by,

\begin{equation}
    \widetilde{W}(\boldsymbol{\epsilon}(\textbf{u}), p_m, d) = \dfrac{1}{2}\left( (1-d)\boldsymbol\epsilon(\textbf{u}) - \dfrac{\alpha}{nK} p_m\mathbb{I}\right) : \mathbb{C} : \left( (1-d)\boldsymbol\epsilon(\textbf{u}) - \dfrac{\alpha}{nK} p_m\mathbb{I}\right),
\end{equation}

which is consistent with an assumption of damage arising in the sub pore scale \cite{chukwudozie2019variational}. Finally, our regularized crack evolution problem is then stated as a minimization principle for the functional $\mathcal{E}_{\ell}(\textbf{u},d,p_m,p_f)$, with respect to the variables $\textbf{u}$ and $d$, with the added condition that the damage process is irreversible:
\begin{equation}\label{variational formulation of phase-field}
    \textbf{u}, d = \underset{\textbf{u},d}{{\operatorname{argmin}}} \ \mathcal{E}_{\ell}(\textbf{u},d,p_m,p_f)\ \text{,\ \   subject to } \dot{d} \ge 0. 
\end{equation}

The following set of evolution equations can then be derived from the Karush–Kuhn–Tucker (KKT) \cite{karush1939minima, kuhn1951nonlinear} conditions:

\begin{equation}\label{basic u problem}
    \nabla \cdot \left( (1-d)^2\ \mathbb{C}:\boldsymbol\epsilon(\textbf{u}) -(1-d)\alpha p_m \mathbb{I}\right) + \textbf{b} = p_f\nabla d, 
\end{equation}

\begin{equation}\label{damage equation}
    S_d \coloneqq  \dfrac{2G_c\ell}{c_0}\Delta d - \dfrac{G_c}{c_0\ell}\zeta'(d)-g'(d)W(\boldsymbol{\epsilon}(\boldsymbol{\textbf{u}})) + \alpha p_m\nabla \cdot \textbf{u} + \nabla \cdot (p_f\textbf{u}) \le 0,
\end{equation}

\begin{equation}
    \dot{d} \ge 0,
\end{equation}

\begin{equation}
    S_d\dot{d} = 0,
    \label{eq:ddot-strong}
\end{equation}

with boundary conditions, $\boldsymbol\sigma \cdot \textbf{n} = \textbf{t}$ on $\partial \Omega_N$ and $(2G_c\ell\nabla d +c_0p_f\textbf{u})\cdot \textbf{n} \ge 0$ on $\partial \Omega$. The damaged stress is defined as $\boldsymbol\sigma = (1-d)^2\ \mathbb{C}:\boldsymbol\epsilon(\textbf{u}) -(1-d)\alpha p_m \mathbb{I}$.

\section{Multi-resolution method }\label{numerics}

\subsection{General overview}

\begin{figure}[!htbp]
    \centering
    \includegraphics[width=\textwidth]{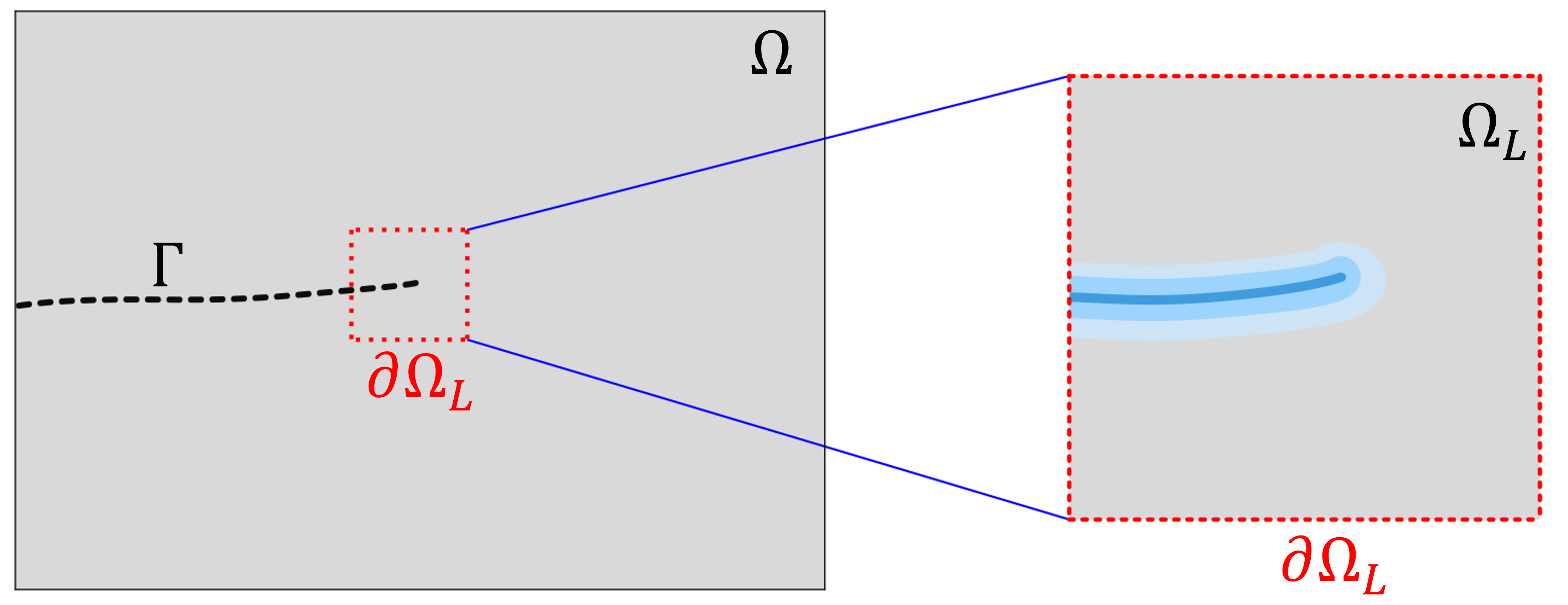}
    \caption{Global domain $\Omega$ on the left and local domain $\Omega_L$ on the right. The surface corresponding to the local domain boundary $\partial\Omega_L$ is indicated within the global domain with the dashed red lines.  The local domain is magnified to highlight the phase-field representation of the global crack $\Gamma$.}
    \label{fig:section_2_figure}
\end{figure}

We propose a multi-resolution method to approximate the solution of the hydraulic fracture problem in porous media. It consists of coupling two problems between a global domain and a local subdomain, as shown in Figure~\ref{fig:section_2_figure}.   In the global domain problem, the governing equations \eqref{linear momentum balance}-\eqref{mass balance fracture} are discretized, and the crack is represented with a sharp geometry.  The crack geometry is assumed to be fixed during a solution step in the global domain, and relevant fields are calculated over the entire domain.  

By contrast, the local subdomain concerns only a portion of the entire domain, namely in the vicinity of crack tips.  It is encapsulated within the global domain as shown in Figure~\ref{fig:section_2_figure}. In the local subdomain, a discretization of the variational principle \eqref{variational formulation of phase-field} is used to simulate crack evolution.  During a solution step in the local subdomain, the pressure fields within the fracture and matrix are assumed to be fixed.  

The two problems are coupled in the following manner.  The displacement and pressure fields are extracted from a solution step in the global domain and passed to the local subproblem in different ways.  In particular, the global displacement fields are extracted along the surface $\partial\Omega_L$. These fields are then  applied as Dirichlet boundary conditions for the local subproblem.  The matrix pressure $p_m$ and pressure in the fracture $p_f$ are transfered as fields from the global domain to the local subdomain (see Section~\ref{sec:pressure_projection} for details), and assumed to be fixed for the local subproblem.   

The aforementioned operations provide everything that is needed for the local subproblem from the global domain.  Based on the imprint of the global crack geometry on the local subdomain, a regularized fracture surface is created (through an initial damage field) and then crack propagation is simulated in the local subdomain. Once an extension of the crack in a scale that can be represented in the global domain is identified, the sharp crack geometry in the global problem is updated accordingly.  

In principle, the aforementioned multi-resolution approach can be implemented using a number of different discretization methods in the global domain and the local subdomain.  In the sections that follow, we describe the particular choices used in this work as well as some important implementation details.  The method is described in a two-dimensional context, but many aspects can be readily extended to three-dimensional problems.  

\subsection{Global problem discretization}
\label{sec:global_disc}

The global problem in our multi-resolution approach encompasses the physics of fluid flow in both the fractures and the pore structure of our domain, as well as the deformation of the solid media. The set of governing equations consists of  \eqref{linear momentum balance}-\eqref{mass balance fracture}, combined with constitutive assumptions \eqref{biot law}-\eqref{poorly compressibility} and appropriate boundary conditions. 

In this work, we use the discretization method proposed by Cusini et al.\ \cite{cusini2021simulation} in the study of fluid flow in fractured porous media. 
The  domain $\Omega$ is partitioned with a  mesh $\mathcal{T}$. Then, the intersection of the fracture network $\Gamma$ with $\mathcal{T}$ defines the fracture triangulation $\mathcal{F}$. These meshes are then used to define discrete counterparts $\textbf{u}^h_G$, $p_m^h$ and $p_f^h$ of the unknown fields $\textbf{u}$, $p_m$ and $p_f$, as well as discrete approximations of equations  \eqref{linear momentum balance}, \eqref{mass balance matrix} and \eqref{mass balance fracture}.

A finite element approximation is constructed for the displacement field and employed in a standard Galerkin approximation to the global force balance \eqref{linear momentum balance}.
The continuous part of the displacement field is approximated with a standard space $\mathcal{U}$  of 4-node bilinear shape functions $\{ {\boldsymbol{\eta}}_a \} $.  In the subset of elements that are ``cut" by the global crack geometry, a space $\mathcal{W}$ of discontinuous enrichment functions $\{\boldsymbol{\phi}_b\}$ is constructed using the formulation described in \cite{linder2007finite}.   

The full displacement field in the global problem is constructed using both continuous and discontinuous parts as
\begin{equation}
    \label{eq:displacement_approx}
    \textbf{u}^h_G = \underbrace{\sum_{a=1}^{n_u}u_a \boldsymbol{\eta}_a}_{\text{continuous part}} + \underbrace{\sum_{b=1}^{n_w}w_b\boldsymbol\phi_b}_{\text{discontinuous part}}, 
\end{equation}
where $\{u_a\}, \{ w_b\}$ are scalar degrees of freedom. 

The flow equations  \eqref{mass balance matrix} and \eqref{mass balance fracture} are discretized with a finite-volume method.   Piecewise-constant pressure fields are constructed for $p_m^h$ and $p_f^h$ over the matrix mesh $\mathcal{T}$ and the fracture mesh $\mathcal{F}$, respectively.  
Fluxes are computed using a two point flux approximation. The interaction between the flow in the fracture and matrix is effected via the embedded discrete fracture model (EDFM) \cite{lee2001hierarchical, hajibeygi2011hierarchical}. 

In terms of the temporal discretization, a backward Euler method is used throughout.  This gives rise to a fully-coupled system of nonlinear equations, whose solution is obtained with a Newton method, in a monolithic fashion. 

One limitation of the construction \eqref{eq:displacement_approx} is that the discontinuous enrichment functions are not capable of representing a crack tip that terminates inside of an element.  As such, any new extension of the crack geometry has to traverse from one side of a new element to another. The implementation of the fracture flow solver requires all cells to have non-zero volumes.  Accordingly, new fracture cells are assigned a  small aperture value ($w_0$). The total discrete aperture of the cell is thus given by $w_h = w_n + w_0$, where $w_n$ is the mechanical aperture that is consistent with the jump provided by the displacement field.  The minimum opening $w_0$ has been interpreted as a representation of the roughness of the fracture surfaces, providing a pathway for fluid flow even when the cracks are mechanically closed. See, for example \cite{cusini2021simulation}.

\subsection{Local problem initialization} 

\subsubsection{Construction of subdomain, submesh and damage-fixed nodes}\label{subdomain_construction}

A local subdomain of size $L\times L$ is constructed by simply centering it on a global crack tip.  The size $L$ is selected to be an even integer multiplier of the global mesh spacing, leading to a square bounding box that conforms to the background mesh.  This choice is adopted for convenience in this work, although other constructions are possible, such as in  \cite{giovanardi2017hybrid}.

To obtain the submesh $\mathcal{T}_L$, we start by considering the restriction of the global triangulation $\mathcal{T}$ to the subdomain $\Omega_L$.  The mesh for the local subdomain is constructed by uniformly refining the set of elements in this restriction. This facilitates the transfer of nodal data from the global to the local problem.  The mesh size $h_{local}$ in the local subdomain is chosen to be sufficiently small to resolve the damage band, of size $\mathcal{O}(\ell)$. In this work, we use  $\ell / h_{local} \approx 4$.

\begin{figure}[h]
    \centering
    \includegraphics[width=\textwidth]{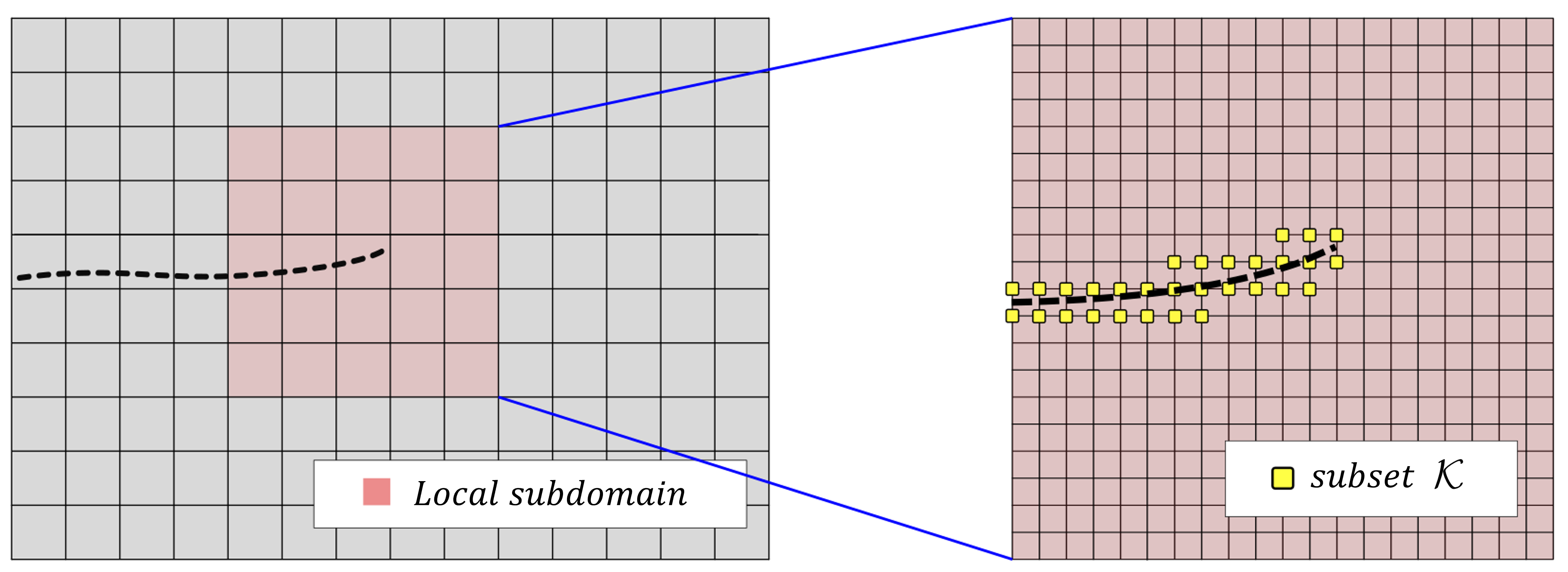}
    \caption{Global domain $\Omega$ on the left and local domain $\Omega_L$ on the right. The nodes in the subset $\mathcal{K}$, where $d$ is set to $1$ are colored in yellow.}
    \label{fig:subset_kappa}
\end{figure} 

The crack is represented by prescribing $d = 1$ in a set $\mathcal{K} \in \mathcal{T}_L$ of nodes in the local subdomain, as shown in Figure \ref{fig:subset_kappa}. This set is constructed in two steps. First, all elements of $\mathcal{T}_L$ which are intersected by the global fracture triangulation $\mathcal{F}$ are identified.  Then, $\mathcal{K}$ is defined to be the set of all nodes that belong to any of these intersected elements.

\subsubsection{Transfer of global pressures to local mesh}
\label{sec:pressure_projection}

Due to the different levels of resolution between the global and local problems, we find it advantageous to transfer the pressure fields in a particular manner.  We note that in the global problem, the fracture pressures $p_f^h$ are available at the cell centers of 1D finite volumes, and the matrix pressures $p_m^h$ are available at the cell centers of 2D finite volumes.  These fields are transferred to quadrature points in the finite element mesh for the local subproblem using the transfer operators 
$\Pi^{\Omega_L}_f$ and $\Pi^{\Omega_L}_m$.

\begin{figure}[!htbp]
    \centering
    \includegraphics[width=6cm]{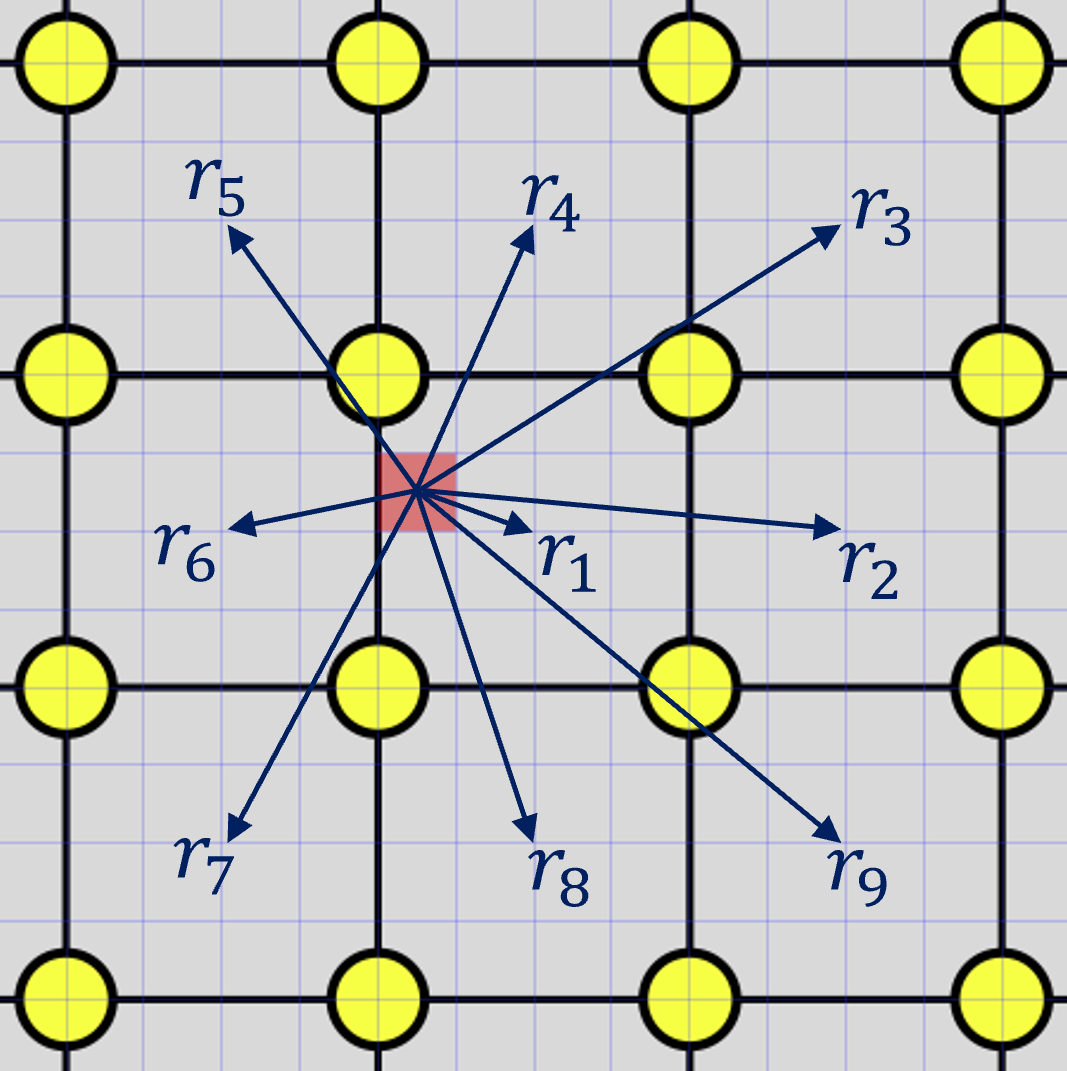}
    \caption{Illustration of global cells (in gray) in the neighborhood of a local element, indicated in pink. The matrix pressure at a quadrature point in the local element is calculated using an average of global pressures from neighboring cells $\{ i \}$, with weights corresponding to the distances $r_i$ between the quadrature point and the cell center.}
    \label{fig:pressure_interpolation}
\end{figure}

The operator $\Pi^{\Omega_L}_f$ that transfers the fracture pressure is straightforward, and inspired by a similar operation described in Santillán et al.\ \cite{santillan2018phase}. Given any point $ {\bf{x}} \in \Omega_L$, and a global fracture pressure field $p_f^h$, $\Pi^{\Omega_L}_fp_f^h({\bf{x}})$ is obtained by finding the closest cell of $\mathcal{F}$ to $x$ and taking the value of $p_f^h$ at this cell. More precisely,

\begin{equation}
    \Pi^{\Omega_L}_f p_f^h({\bf{x}}) = p_f^h(\underset{c \in \mathcal{F}}{\text{arg min }}\text{dist}(c,{\bf{x}} )).
\end{equation}

For the operator $\Pi^{\Omega_L}_m$ that transfers the matrix pressure field, we use an averaging procedure.  This has the effect of smoothing the matrix pressure field at the resolution of the local mesh.  At a quadrature point ${\bf{x}}_Q \in \Omega_L$, the matrix pressure in the local domain is obtained from a weighted average of the pressures in the global cells that surround the point.  Specifically, 
\begin{equation}
    \Pi^{\Omega_L}_m p_m^h({\bf{x}}_Q) = \dfrac{\sum_{i=1}^s r^{-1}_ip_m^h(c_i)}{\sum_{i=1}^s r^{-1}_i},
\end{equation}
where $r_i$ denotes the distance between the quadrature point location and the center of global cell $i$, as indicated in Figure~\ref{fig:pressure_interpolation}.  In the sum, all cells that neighbor the global cell containing quadrature point ${\bf{x}}_Q$ are used.  In the particular case when a quadrature point happens to reside in the center of the local element and $r_1 = 0$, the above sum is replaced with   
$ \Pi^{\Omega_L}_mp_m^h({\bf{x}}_Q) = p^h_m (c_1)$.

\subsection{Local problem discretization}

With the local subdomain properly identified and initialized, we now describe the additional steps to discretize the displacement and damage fields in the local subdomain and solve for their approximations.
The governing equations for the macro-force balance \eqref{basic u problem} and the damage evolution \eqref{damage equation} are both treated with the finite element method.  The damage and displacement fields are both approximated using four-node bilinear quadrilateral elements. 

Let $\Omega_L$ denote the local domain. From the finite-element approximation $\textbf{u}_G^h$ computed in the global problem, we extract $\textbf{u}^h_G|_{\partial\Omega_L}$ and use it to constrain the displacements on the boundary of the subproblem \footnote{In the multi-resolution method of Muixi et al.\ \cite{muixi2021combined},  the displacement boundary conditions near the crack base were released. We did not find this to be necessary in our approach.}.  As such, the trial space $\boldsymbol{\mathcal{U}}^h$ is given by
\begin{equation}\label{disp subspace}
    \boldsymbol{\mathcal{U}}^h = \{ \textbf{u}_L^h \in H^1(\Omega_L)^n \mid \textbf{u}_L^h = \textbf{u}^h_G \text{ on } \partial\Omega_L \}.
\end{equation}

The function space $\mathcal{D}^h$ of admissible damage fields is given by 
\begin{equation}\label{damage subspace}
    \mathcal{D}^h = \{ d^h \in H^1(\Omega_L) \mid d^h = 1 \text{ on } \mathcal{K} \},
\end{equation}
where $\mathcal{K}$ denotes the set of damage-fixed nodes that correspond to the global crack at the beginning of the load step, as described in subsection~\ref{subdomain_construction}. 

The test spaces are given by $\boldsymbol{\mathcal{W}}^h = H_0^1(\Omega_L)^n$ for the displacements and $\mathcal{C}^h = \{ c^h \in H^1(\Omega_L) \mid c^h = 0 \text{ on } \mathcal{K} \}$ for the damage field. The Galerkin approximation to the problem on the local subdomain is then:
\medskip

\begin{mdframed}[
    frametitle={The spatially discrete form},
    frametitlebackgroundcolor=gray!20,
    backgroundcolor=gray!5,
    linewidth=0pt,
    nobreak=true
  ]
  Find $\textbf{u}_L^h \in \boldsymbol{\mathcal{U}}^h$ and $d^h \in \mathcal{D}^h$, such that $\forall \textbf{w}^h \in \boldsymbol{\mathcal{W}}^h$ and $\forall c^h \in \mathcal{C}^h$,
 
 \begin{align}
  \left( \nabla \textbf{w}^h, \boldsymbol\sigma_L^h \right) - \left( \textbf{w}^h, \Pi^{\Omega_L}_f p_f^h \nabla d^h \right) - \left( \textbf{w}^h, \textbf{b} \right) - \left< \textbf{w}^h, \textbf{t} \right>_{\partial_t\Omega} &= 0, \label{eq: semidiscrete momentum balance}   \\
  \begin{split}
    \frac{2G_c^h\ell}{c_0}\left( \nabla c^h, \nabla d^h \right) + \frac{G_c^h}{c_0\ell}\left( c^h, \zeta'(d^h) \right) + \left( c^h, 2(d^h-1)\Psi^{<A>}(\textbf{u}_L^h) \right) \\
    - \left( c^h, \alpha \Pi^{\Omega_L}_m p_m^h\nabla \cdot \textbf{u}_L^h \right) + \left( \nabla c^h, \Pi^{\Omega_L}_f p_f^h\textbf{u}_L^h \right)  &= 0, \label{eq: semidiscrete damage evolution}    
  \end{split} 
\end{align}
    
\end{mdframed}
where we have used $(\cdot,\cdot)$ to denote the standard inner product in the $L^2(\Omega)$ space, and $\left< \cdot, \cdot \right>$ to denote its restriction on the boundary.   In the above,  $\boldsymbol\sigma^h_L$ denotes the discrete stress, given by 
\begin{equation}
    \boldsymbol\sigma^h_L = \left( (1-d^h)^2\ \mathbb{C}:\boldsymbol\epsilon(\textbf{u}_L^h) -(1-d^h)\alpha p_m^h \mathbb{I}\right).
\end{equation}

The discrete active strain energy $\Psi^{<A>} ({\bf{u}}^h_L)$ that serves as a driver for the damage evolution can be constructed from the total strain energy $\Psi(\textbf{u}) = (1/2)\boldsymbol\epsilon(\textbf{u}^h):\mathbb{C}:\boldsymbol\epsilon(\textbf{u}^h)$ in a number of ways.  Options include  the spectral split \cite{miehe2010phase} or the volumetric-deviatoric split \cite{amor2009regularized}. We employ the spectral split in this work, unless otherwise indicated.
 
We employ an alternating minimization scheme to solve the coupled system of equations \eqref{eq: semidiscrete momentum balance}-\eqref{eq: semidiscrete damage evolution}. Convergence is measured with respect to the $L_2$-norms of the change in the damage $d^h$ and  displacement fields $\textbf{u}_L^h$, using a relative tolerance of $10^{-4}$. With this approach, each of the equations becomes linear with respect to its primary variable, simplifying the solution process.

We note that \eqref{eq: semidiscrete damage evolution} does not explicitly include the irreversibility constraint  \eqref{eq:ddot-strong}.  We have found this to be unnecessary in the multi-resolution scheme, as the Dirichlet condition $d^h=1$ on the crack set $\mathcal{K}$ is sufficient to prevent any healing of the fracture surface relative to the global crack. 
In terms of the local dissipation function $\zeta(d)$, in this work we use 
$\zeta(d) = d$ which corresponds to the AT-1 phase-field model of fracture.  This is selected due to the fact that it gives rise to a compactly supported damage field and a fully elastic stage prior to damage initiation \cite{pham2011gradient}.

Finally, we note that phase-field models of fracture tend to give rise to a mesh and regularization length dependent critical fracture energy that is larger than $G_c$ \cite{bourdin2008variational}.  To account for this, we use a discrete value of $G_c^h$ that is obtained as a function of the local mesh spacing and regularization length, as
    \begin{equation}
        G_c^h = G_c \left( 1 + \dfrac{ h_{local} }{c_0\ell} \right)^{-1},
    \end{equation}
such that the effective critical fracture energy is very close to that of the material.  

\subsection{Crack propagation: translating local damage updates into global crack extensions}\label{propagation_step}

A key step of the algorithm concerns how changes to the damage in the local subdomain are translated into updates to the crack geometry in the global domain.  We provide details of our algorithm for this procedure here.  The scheme is presented in the context of a single crack tip that is propagating through the global domain, but the generalization to multiple cracks is straightforward.

\begin{figure}[!h]
\centering
\begin{subfigure}{.33\textwidth}
  \centering
  \includegraphics[width=\linewidth]{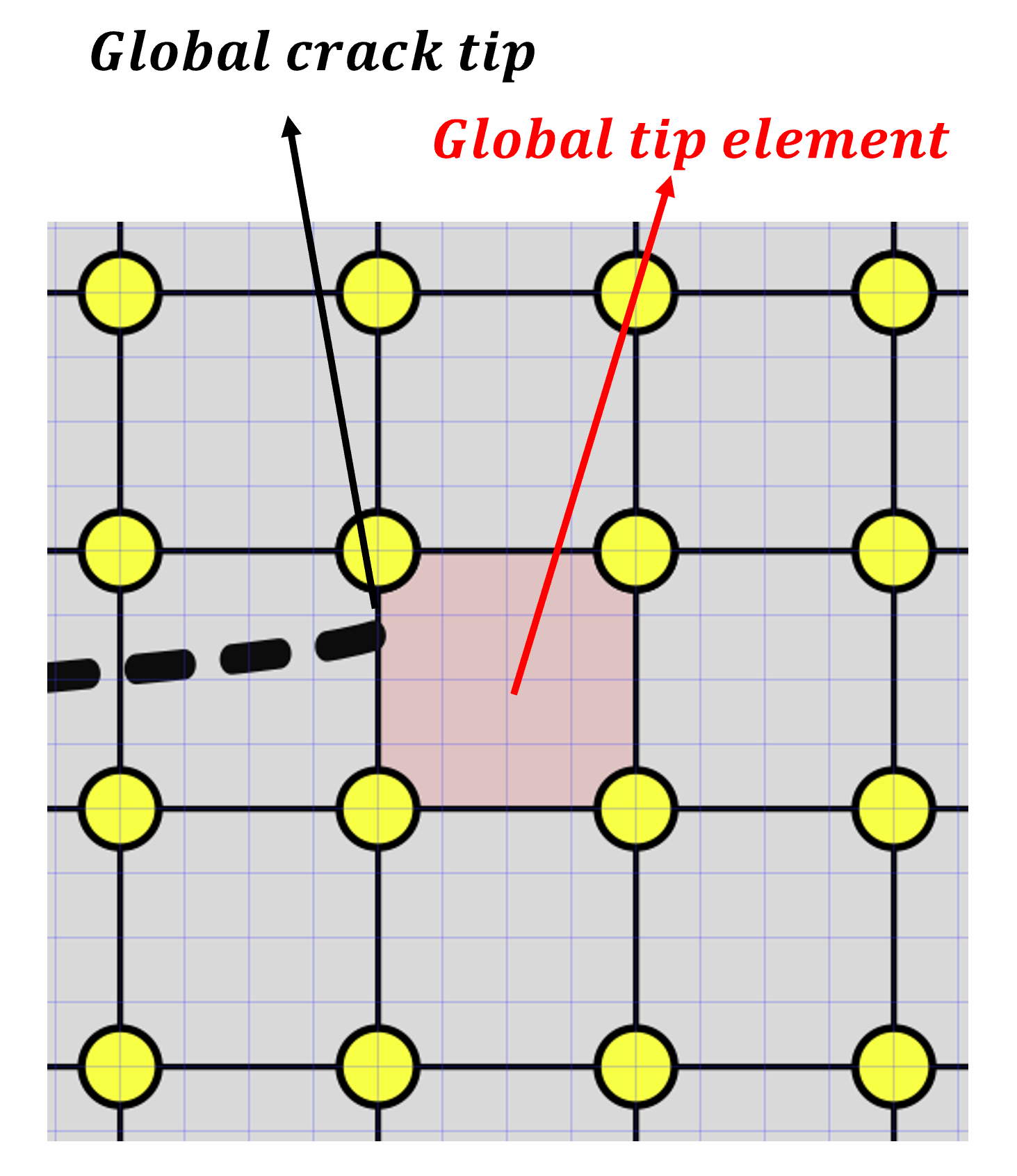}
  \caption{}
  \label{fig:prop_items}
\end{subfigure}%
\begin{subfigure}{.33\textwidth}
  \centering
  \includegraphics[width=\linewidth]{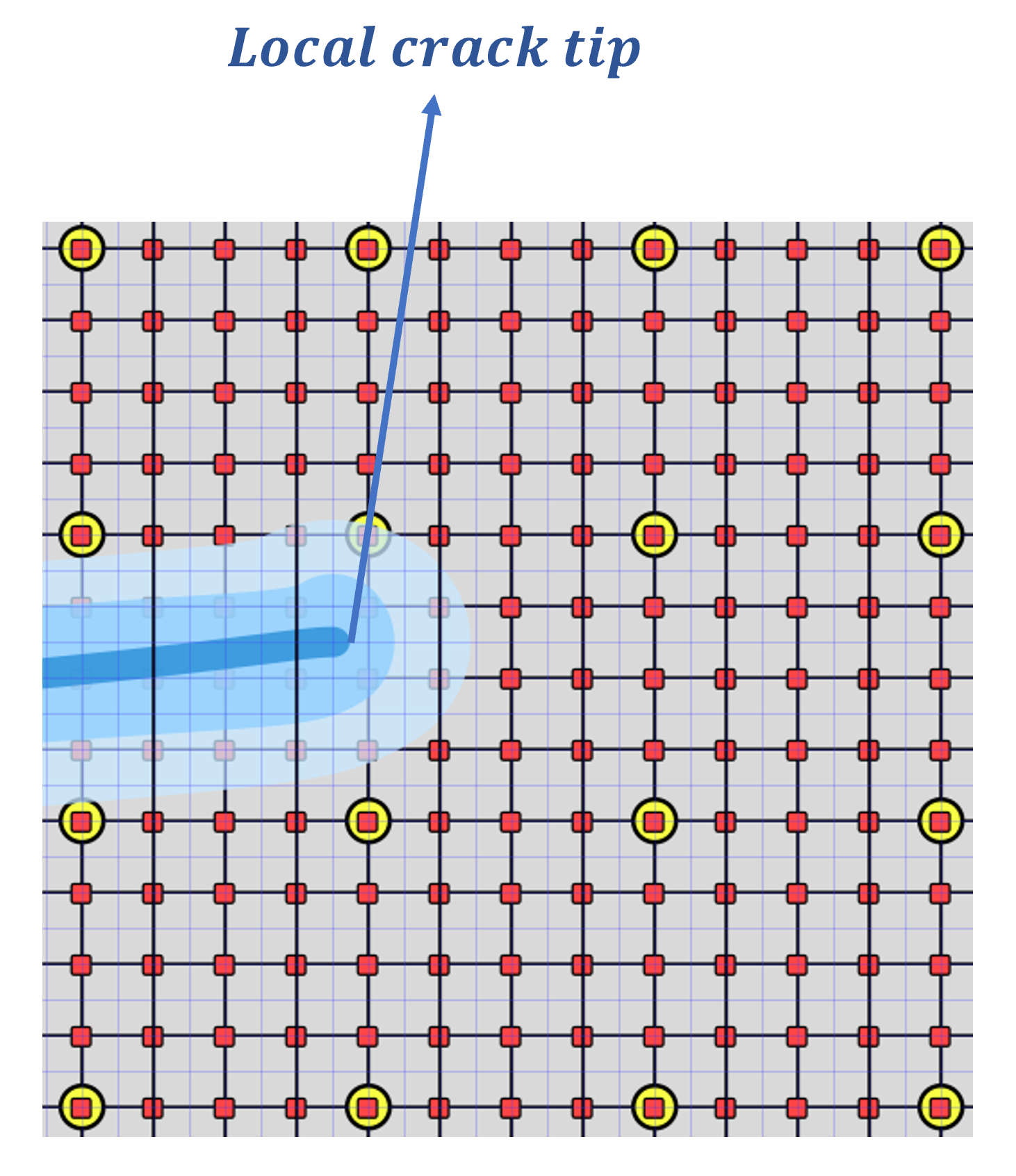}
  \caption{}
  \label{fig:prop_items2}
\end{subfigure}
\begin{subfigure}{.33\textwidth}
  \centering
  \includegraphics[width=\linewidth]{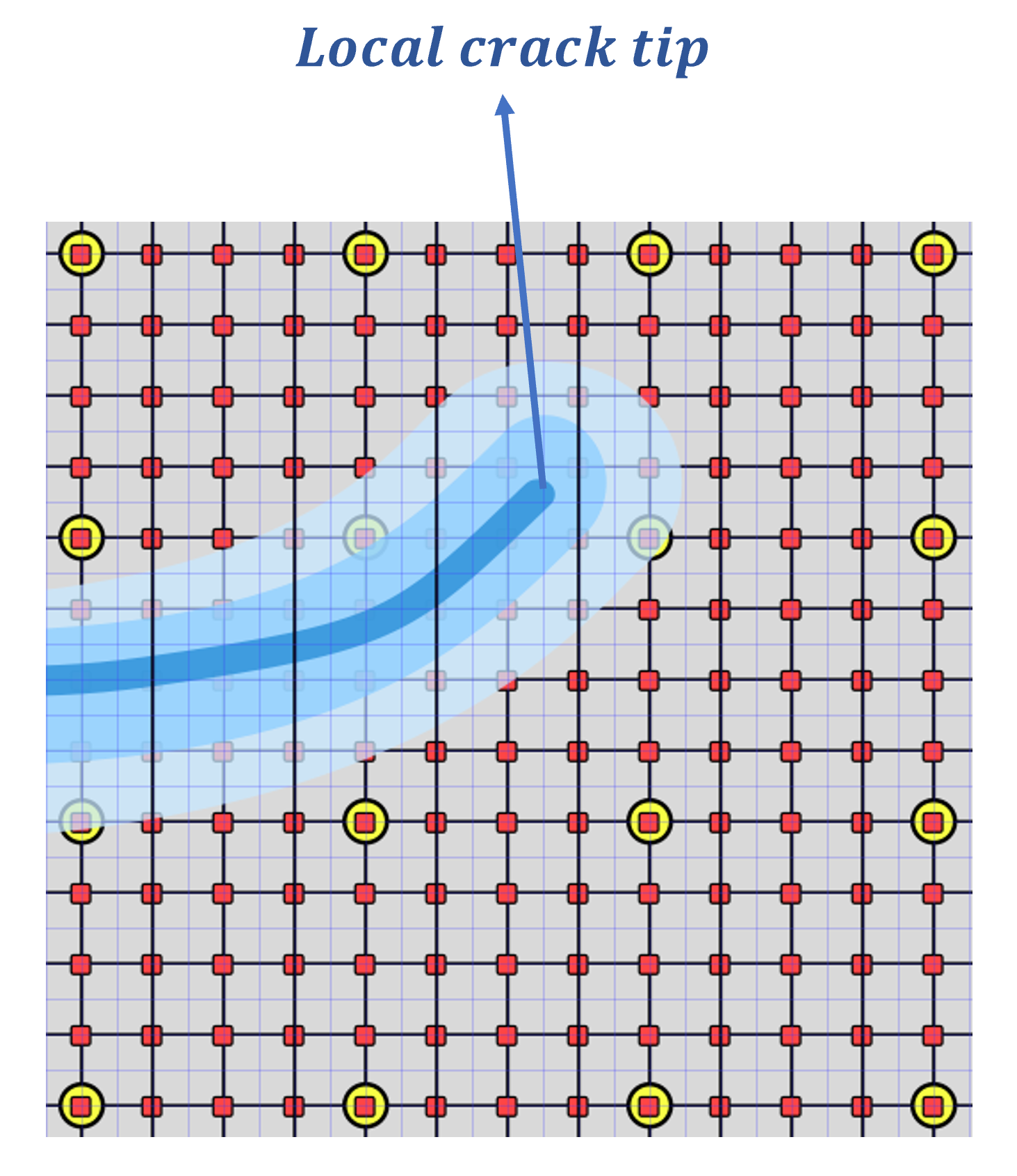}
  \caption{}
  \label{fig:prop_items3}
\end{subfigure}
\caption{(a): Illustration of the global crack tip and global tip element. (b) Case when the local tip falls inside the tip element. (c) Case when the local tip falls outside the tip element.}
  \label{fig:all_tips}
\end{figure}

At each step in the solution algorithm, in the global problem, we keep track of two geometric entities, namely the global crack tip and the global tip element (Figure \ref{fig:all_tips}a). The global crack tip is defined by the interior endpoint of the crack.  The global tip element is the element that contains the tip on one of its sides, but is not yet ``cut" by the crack.  In essence, it is the global element just ahead of the propagating crack tip.   In the local subproblem, we identify a local version of the crack tip that is obtained from the discrete damage field $d^h$ (Figure \ref{fig:all_tips}b and \ref{fig:all_tips}c).  
The algorithm to extend the global crack geometry depends on the relative location of the global crack tip, global tip element, and local crack tip, as described below.

\begin{figure}
\centering
\begin{subfigure}{.331\textwidth}
  \centering
  \includegraphics[width=.99\linewidth]{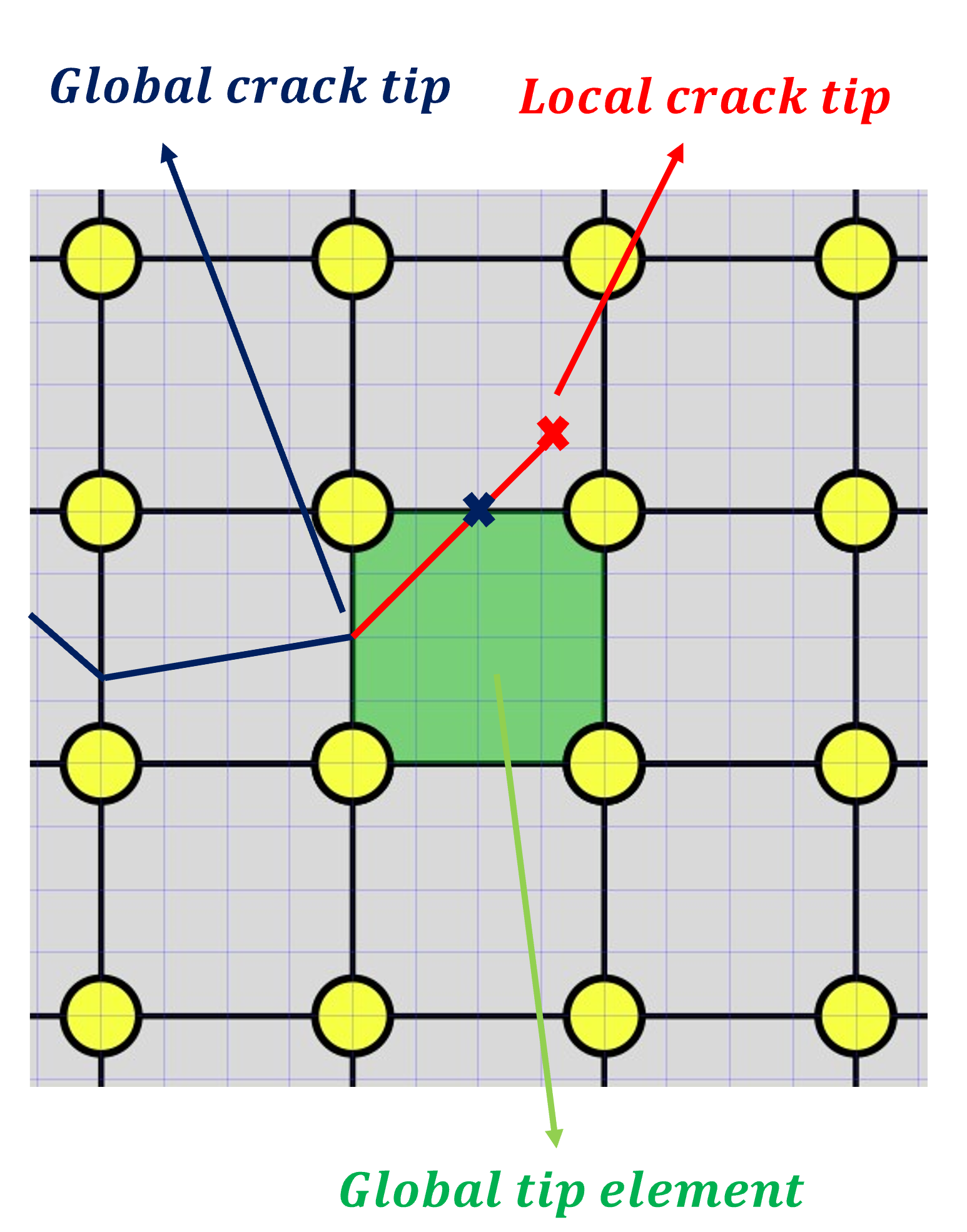}
  \caption{}
  \label{fig:prop_1}
\end{subfigure}%
\begin{subfigure}{.33\textwidth}
  \centering
  \includegraphics[width=\linewidth]{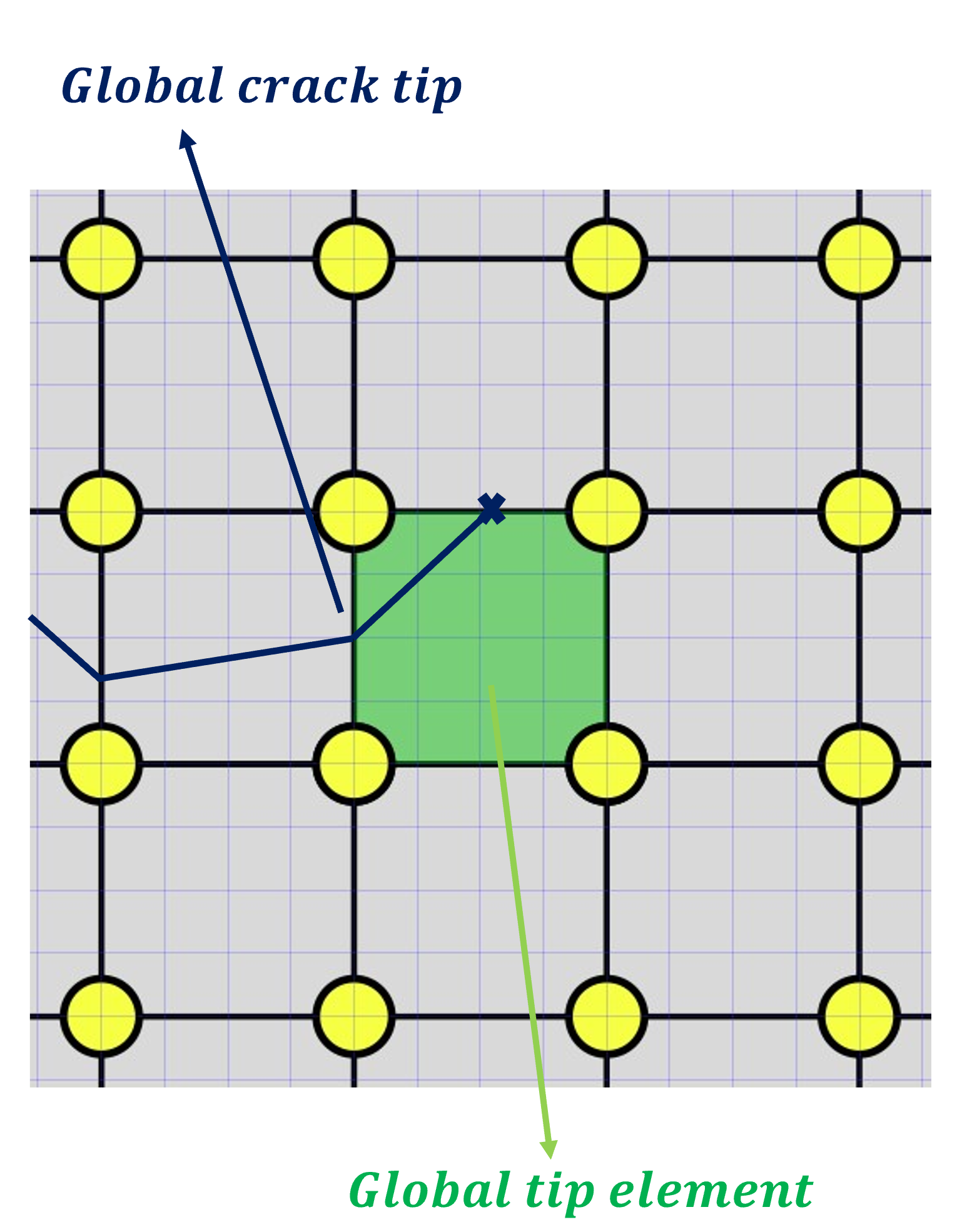}
  \caption{}
  \label{fig:prop_2}
\end{subfigure}
\begin{subfigure}{.33\textwidth}
  \centering
  \includegraphics[width=\linewidth]{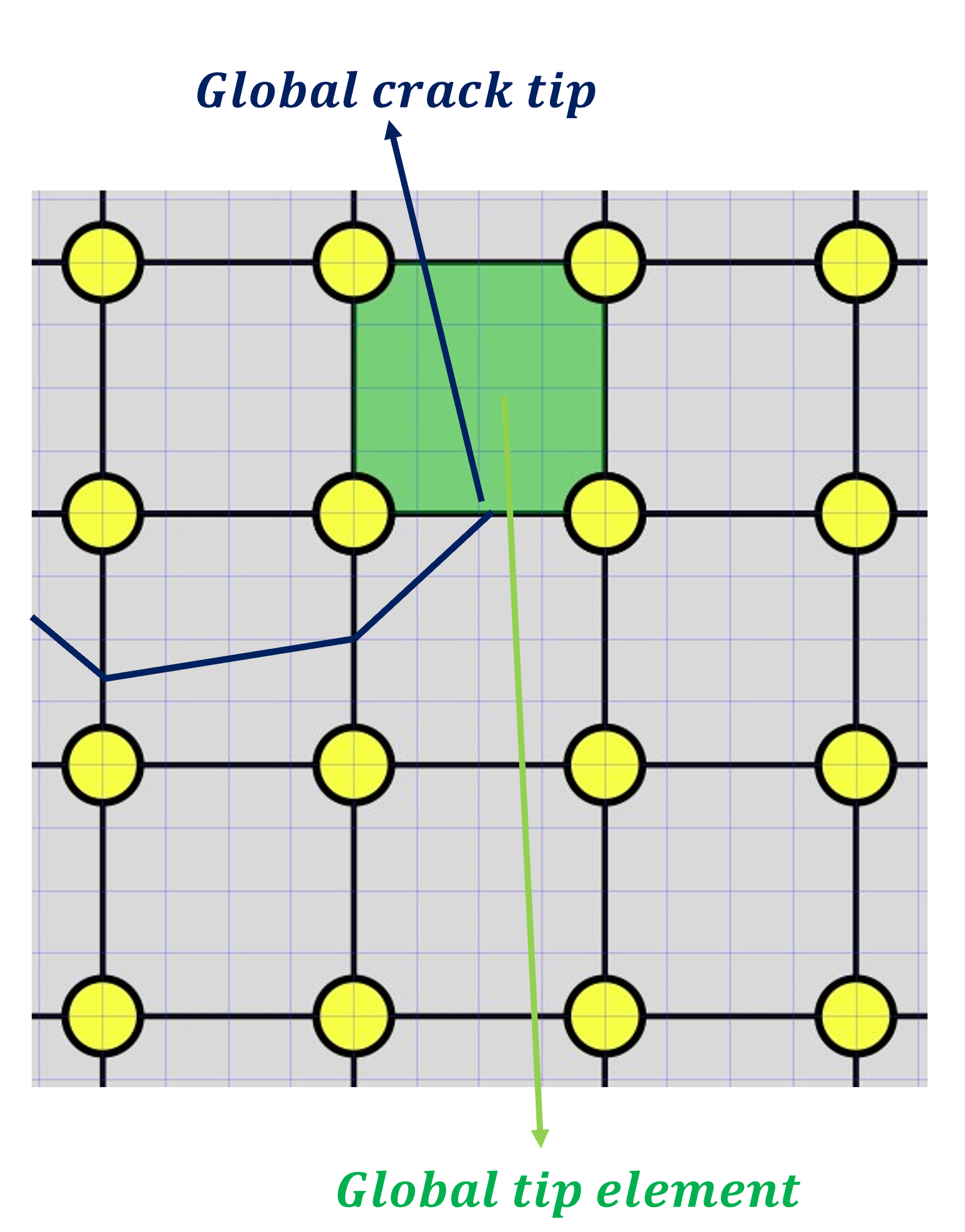}
  \caption{}
  \label{fig:prop_3}
\end{subfigure}
\caption{(a): Construction of the segment connecting global and local tips.  (b) New crack segment to be added. (c) Update of the global crack tip and global tip element.}
  \label{fig:tip_progression}
\end{figure}

The enrichment strategy described in Section~\ref{sec:global_disc} requires global elements that are completely cut by the crack geometry.  As such, any extension of the crack geometry at the global scale must correspond to the global tip element being fractured.  Accordingly, global crack propagation is triggered whenever the local tip falls outside the tip element. In this case, the new global tip is identified by connecting the current global tip and the local tip. Since the local tip is outside the tip element, a new segment will intersect the perimeter of the tip element exactly once (neglecting the obvious intersection at the current global tip). This process is illustrated in Figure \ref{fig:tip_progression}.

Any crack advance beyond this new tip location is neglected at this point, and the algorithm returns to a new global solve with an updated crack geometry.  

In general, the problem of extracting a sharp crack front from a diffuse representation is not trivial. When the crack evolution involves complex topological changes, it is particularly difficult\cite{tamayo2015medial}. In the numerical examples studied in this manuscript, we take advantage of the relatively simple fracture geometry and predicted crack patterns to simplify this process.  In particular, we first identify all elements in the local subdomain with nodes whose damage values are all above a threshold $d_{tr}$. A similar approach was proposed in \cite{giovanardi2017hybrid}. The local crack tip is taken to be the center of the element in this set that is farthest from the base of the crack in the local subdomain.  The threshold used for this process is taken to be $d_{tr} = 1 - h_{local}/(2\ell)$, which is based on the estimate for a damage field near a crack tip given in \cite{yoshioka2020crack}. In essence, for a phase-field model of fracture, this threshold identifies nodes that are expected to correspond to the peaks of the discrete damage field.  For a damage band resolved with a mesh spacing of $\ell / h_{local} = 4$, this gives rise to a threshold of  $d_{tr} = 0.875$. 

\subsection{Algorithm summary}

\begin{figure}[h]
    \centering
    \includegraphics[width=16.5cm]{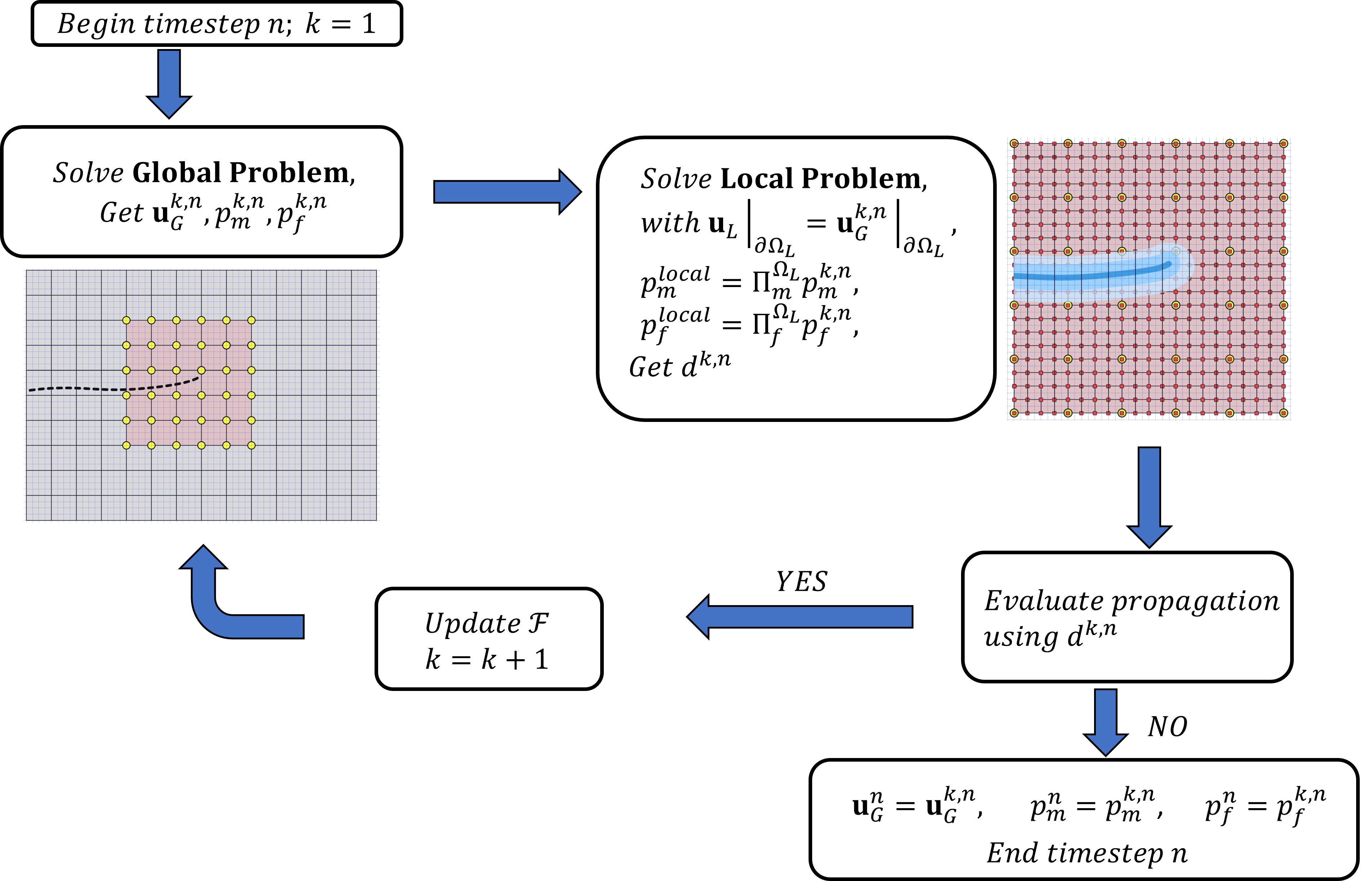}
    \caption{Multi-resolution solution algorithm.}
    \label{fig:solution_algorithm}
\end{figure}

Having described the solution strategies for both the global and local problems and the transfer of various quantities, we now detail the algorithm \eqref{mr_algorithm} that couples the two problems together to simulate crack propagation. Figure~\ref{fig:solution_algorithm} provides an illustration of the algorithm. Within each time step (outer loop), the algorithm employs an inner loop that allows the global problem to be updated as soon as any large enough change in the crack geometry is detected in the local subproblem.  The inner loop is terminated when the propagation step, described in subsection \ref{propagation_step} does not identify any crack advance.  The construction with two nested loops can be viewed as an implicit treatment of the fracture front position, which, according to Lecampion et al.\ \cite{lecampion2018numerical} tends to be more accurate and robust, permitting the use of larger time steps.

\medskip

\begin{algorithm}[H]\label{mr_algorithm}
\small
\SetAlgoLined
 Define initial and boundary conditions
 
 $n = 1$ \\
 $n_F = endStep$
 
 \While{$n \le n_F$}{
 
    \medskip
    
    $\mathcal{F}_n^{1} = \mathcal{F}_{n-1}$
 
    \For{$1 \le k \le maxIter\footnote{The index $k$ is only a dummy variable for this loop that searches for the correct fracture geometry at a given time step.}$}{
    
    \medskip
    
    (1) Solve \textbf{Global Problem} and obtain $\textbf{u}_G^{k,n}, p_m^{k,n}, p_f^{k,n}$

    \medskip
    
    (2) Construct subdomain $\Omega_L$ and submesh $\mathcal{T}_L$. Identify subset of cracked nodes $\mathcal{K}$.
    
    \medskip
    
    (3) Prescribe local boundary conditions $\textbf{u}^h_{L}|_{\partial\Omega_L} = \textbf{u}_G^{k,n}|_{\partial\Omega_L}$ and $d^h = 1$ on $\mathcal{K}$.
    
    \medskip
    
    (4) Construct local pressure fields $p_m= \Pi^{\Omega_L}_m (p_m^{k,n})$ and $p_f = \Pi^{\Omega_L}_f (p_f^{k,n})$
    
    \medskip

    (5) Solve the \textbf{Local Problem} to obtain $d^{k}$.
    
    
    \medskip
    
    
    (6) Use $d^{k}$ and the propagation step (subsection \ref{propagation_step}) to update discrete fracture $\mathcal{F}_n^{k}$.
    
    
    \medskip
    
    \If {$\mathcal{F}_n^{k+1}$ = $\mathcal{F}_n^{k}$}{
    
        \medskip
    
        n = n + 1
        
        \medskip
        
        $\mathcal{F}_n = \mathcal{F}_n^{k}$
        
        \medskip
            
        $\textbf{u}_G^n = \textbf{u}_G^{k,n}$
        
        \medskip
            
        $p_m^n = p_m^{k,n}$
        
        \medskip
        
        $p_f^n = p_f^{k,n}$
        
        \medskip
            
        break
        
        \medskip
    
        }

    }
    
 }
 \caption{Solution algorithm for multi-resolution hydraulic fracture}
\end{algorithm}

\newpage 

\section{Numerical Results}\label{results_section}

We now present the results from various benchmark problems in fluid-driven fracture propagation.  The problems range from those in which flow is only present within the cracks to fully coupled problems involving flow in both the matrix and the evolving manifold that is the fracture geometry.  

\begin{figure}[!htbp]
\centering
\begin{subfigure}{.5\textwidth}
  \centering
  \includegraphics[width=.79\linewidth]{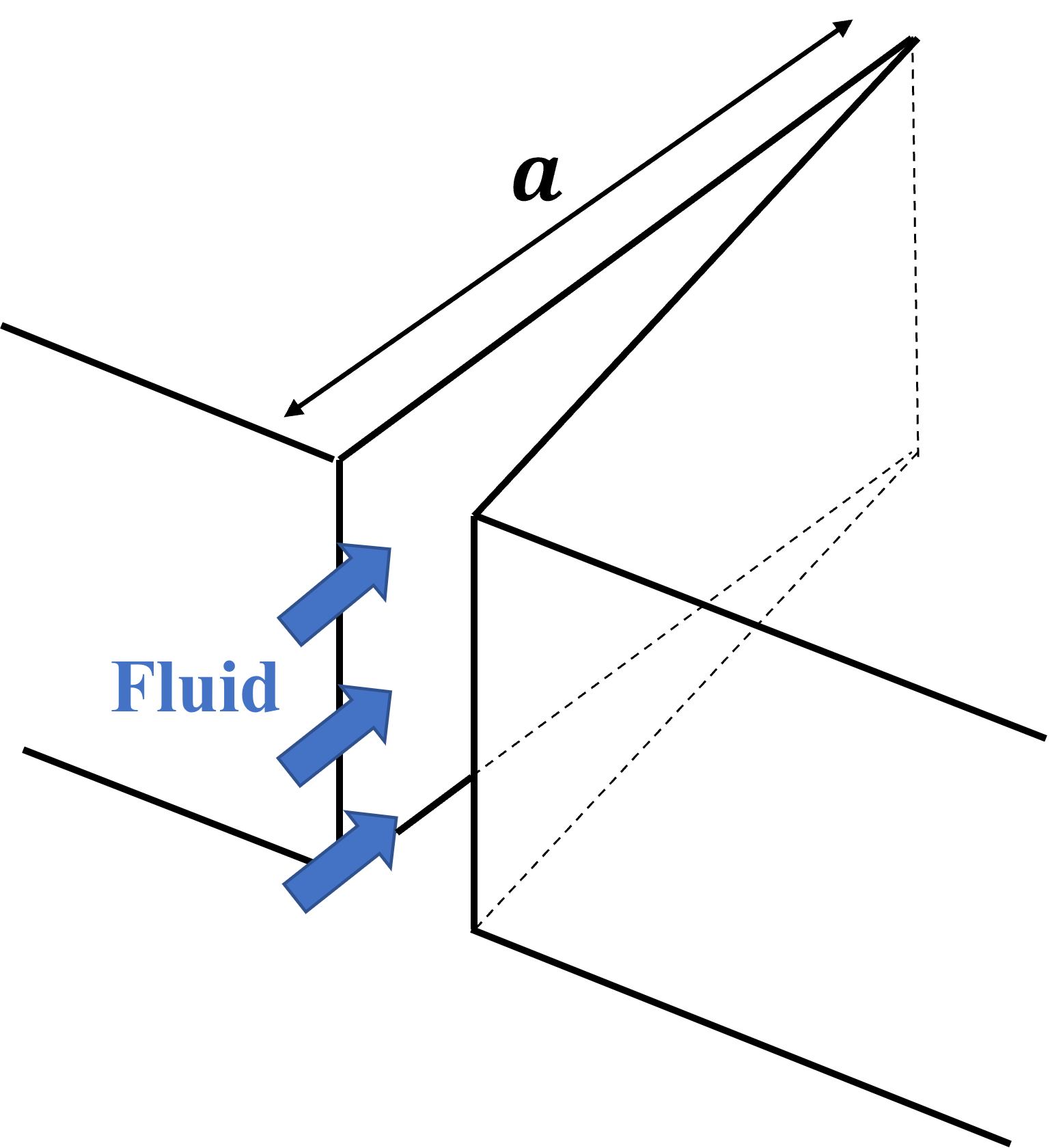}
    \caption{}
    \label{fig:3D_schematics_kgd}
\end{subfigure}%
\begin{subfigure}{.5\textwidth}
  \centering
  \includegraphics[width=.62\linewidth]{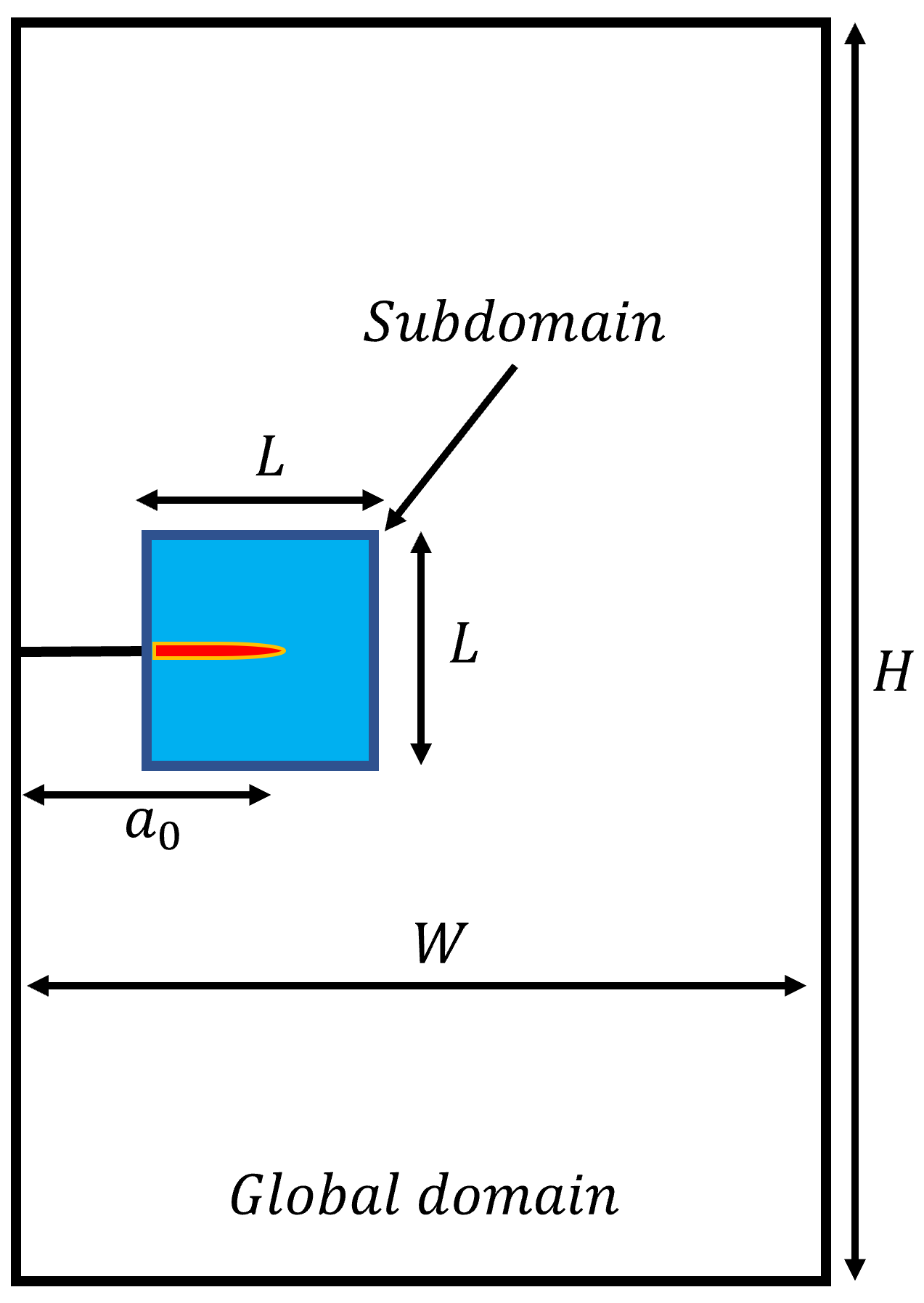}
  \caption{}
  \label{fig:2D_schematics_kgd}
\end{subfigure}
\caption{(a) Schematic of the KGD problem, inspired by \cite{adachi2007computer}. (b) Computational domain for the KGD problem (not to scale).}
  \label{fig:kgd_schematics}
\end{figure}

\subsection{KGD problems} 
We begin with the well-known Khristianovic, Geertsma and De Klerk (KGD) problems of hydraulic fracture \cite{geertsma1969rapid, zheltov19553}.  The problems concern the propagation of a planar fracture in an infinite, impermeable domain, under the injection of a viscous fluid at a constant rate (Figure \ref{fig:kgd_schematics}). 

The response of the system can be characterized by the dimensionless group $\mathcal{K}$ \cite{detournay2016mechanics}:
\begin{equation}\label{KGD_group}
    \mathcal{K} = \dfrac{4G^{1/2}_{c}}{(6\pi^2E'Q\mu)^{1/4}},
\end{equation}
where $E' = E/(1-\nu^2)$. Values of $\mathcal{K} > 4$ correspond to the ``toughness dominated regime" in which crack growth is largely controlled by the fracture toughness of the media.  By contrast, when  $\mathcal{K} < 1$, the fracture toughness is relatively small and the fluid viscosity is the main factor controlling the speed of crack growth.  In the following, we explore the performance of the multi-resolution algorithm to simulate problems in both regimes.  

\subsubsection{Toughness-dominated regime}
The toughness dominated regime is characterized by the creation of new fracture surfaces accounting for almost all of the energy dissipation. 
In this scenario, the fluid can be considered to be inviscid, which leads to a constant pressure distribution over the entire crack.  The assumption of inviscid flow reduces much of the complexity, allowing for the construction of a simple analytical solution for this problem. 
Here, we consider a problem in the toughness-dominated regime resulting from the material parameters and settings given in Table \ref{parameters_tKGD}.  Using these values in \eqref{KGD_group}, we obtain $\mathcal{K} = 6.54$. 

\begin{table}[ht]
\centering
\caption{Material properties and problem parameters for the toughness-dominated KGD problem}
\begin{tabular}[t]{lcc}
\hline
&Value &Unit \\
\hline
Young's modulus ($E$)&16.0&GPa\\
Poisson's ratio ($\nu$)&0.18&--\\
Fluid viscosity ($\mu$)&1.0$\times10^{-12}$&$\text{GPa . s}$\\
Energy release rate ($G_c$)&1.85$\times10^{3}$&$\text{N/m}$\\
Injection rate ($Q$)&1.0$\times10^{-3}$&$\text{m}^2/\text{s}$\\
Initial crack size ($a_0$) &4&$\text{m}$\\
\hline
\end{tabular}
\label{parameters_tKGD}
\end{table}%

The analytical solution for this problem can be separated into two stages as a function of time. 
In the first stage, the pressure builds linearly with time and the crack does not propagate, as the pressure is below the critical threshold $p_{cr} =(G_cE'/\pi a_0)^{1/2}$.  The pressure then reaches the critical value at $t=t_{cr}$, after which the crack begins to propagate. In the second stage the propagation is stable, since the amount of fluid injected is finite and crack propagation leads to a pressure drop as the total space available for the fluid to occupy increases.

The solution for the crack length and pressure in the crack can be written as

\begin{equation}\label{length_solution_tkgd}
    a(t) =     \begin{cases}
      a_0, &  t \le t_{cr},  \\
      \left(\dfrac{E'(Qt)^2}{\pi G_c}\right)^{1/3}, &  t \ge t_{cr}, 
    \end{cases}
\end{equation}

\begin{equation}\label{pressure_solution_tkgd}
    p(t) =     \begin{cases}
      \dfrac{t}{t_{cr}}p_{cr}, &  \ t \le t_{cr},\\
      \left(\dfrac{E'G_c^2}{\pi Qt}\right)^{1/3}, &  t \ge t_{cr},
    \end{cases}
\end{equation}

where $t_{cr} =(\pi G_c a_0^3/Q^2E')^{1/2}$. A derivation of this solution can be found in Yoshioka \cite{yoshioka2020crack}.

For simulations using the multi-resolution scheme, a computational domain of size $W\times H = 30\text{ m} \times 240\text{ m}$ is used (Figure \ref{fig:2D_schematics_kgd}). According to \cite{isida1973analysis}, this domain size should be sufficiently large to yield a good approximation to an infinite plane.  We first report results using a sub-domain size of $L \times L = a_0 \times a_0$.   The sensitivity of the results to the choice of sub-domain size will be discussed later in this subsection.  

In what follows, we report results from a refinement study.  In particular, we report results for three different regularization lengths of decreasing value, beginning with $\ell = a_0/15$.  As the regularization length is decreased, we maintain a mesh size in the subdomain of $h_{local} = \ell/{4}$.  This allows the finite-element approximation to sufficiently resolve the regularized fracture band.  We also note that the effective fracture toughness in the computational problem depends on $h/\ell$ \cite{yoshioka2020crack}.  In the global domain, we maintain the mesh spacing at the ratio of $h_{global} = 3h_{local}$.  For the coarsest global mesh, this translates into roughly 20 elements over the span of the initial crack geometry.  This level of resolution is consistent with those found to be sufficient for resolving pressurized cracks using the Embedded Finite Element Method, in Cusini et al.\ \cite{cusini2021simulation}.

\begin{figure}
\centering
\begin{subfigure}{.5\textwidth}
  \centering
  \includegraphics[width=.99\linewidth]{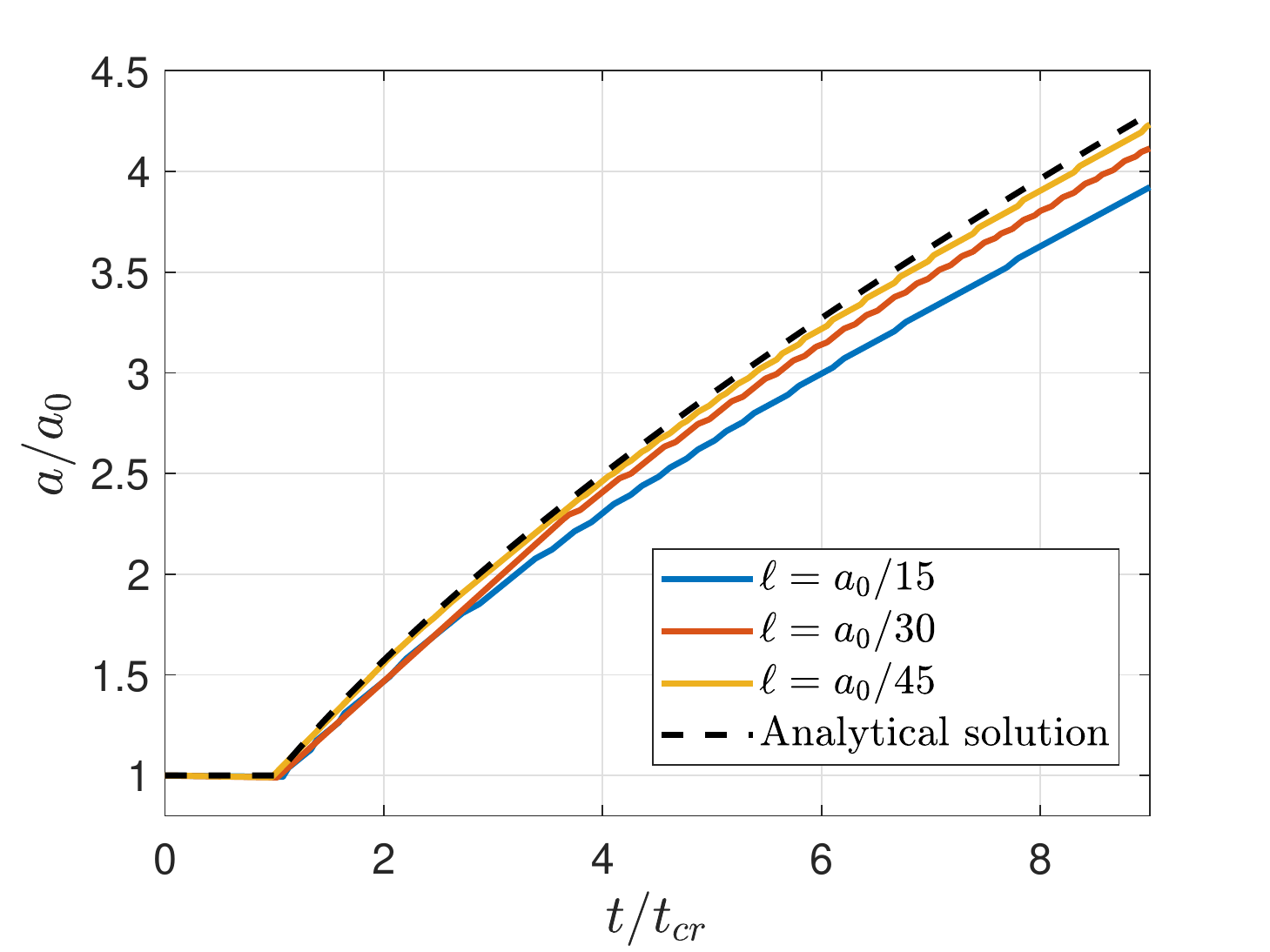}
  \caption{}
  \label{fig:tkgd_length}
\end{subfigure}%
\begin{subfigure}{.5\textwidth}
  \centering
  \includegraphics[width=.99\linewidth]{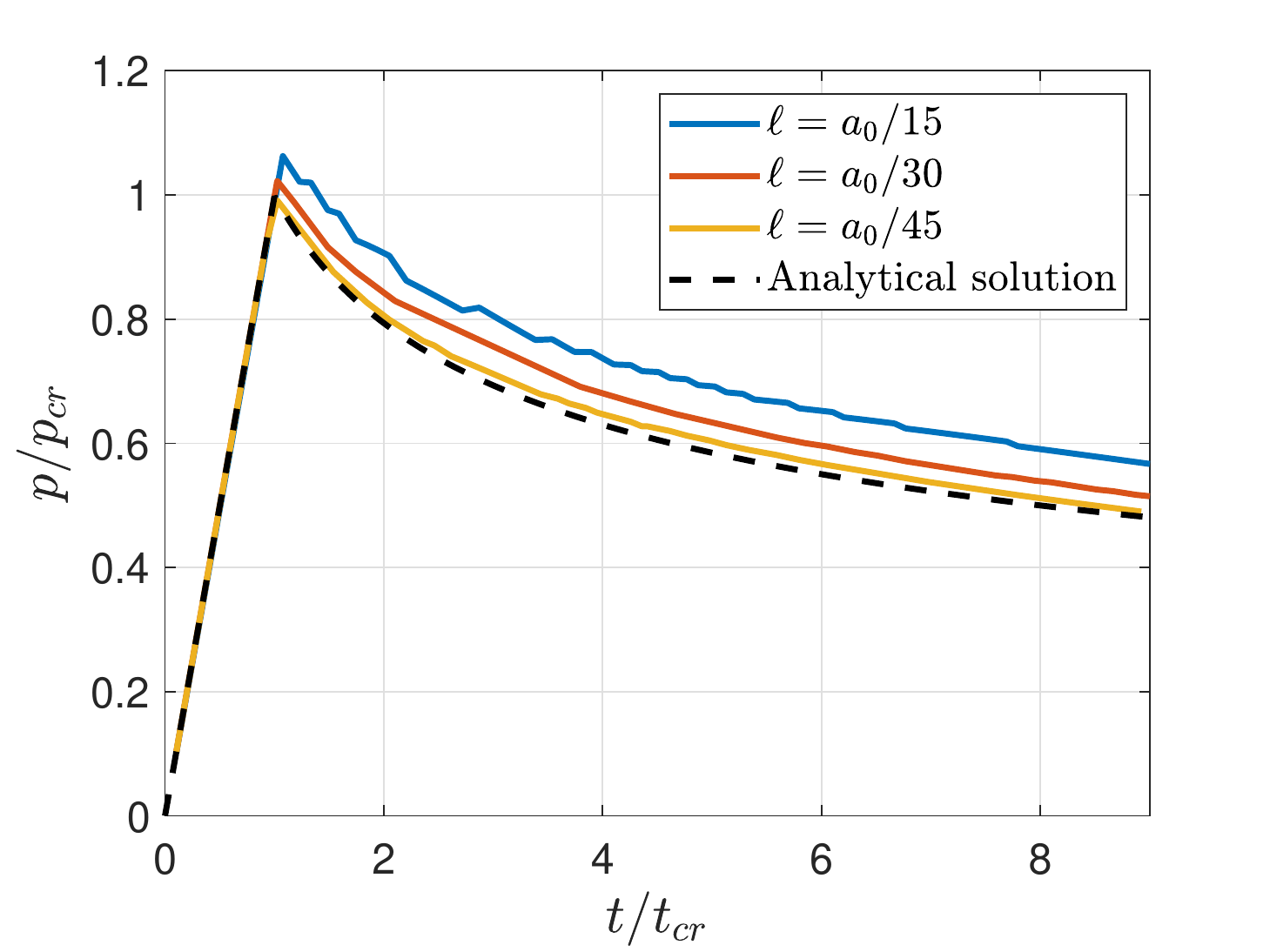}
  \caption{}
  \label{fig:tkgd_pressure}
\end{subfigure}
\caption{Comparison of numerical results and analytical solution for the toughness-dominated KGD problem: (a) crack length vs.\ time and (b) pressure vs.\ time.}
  \label{fig:tkgd_charts}
\end{figure}

The results obtained with the multi-resolution scheme are compared to the analytical solutions \ref{length_solution_tkgd} and \ref{pressure_solution_tkgd} in Figures \ref{fig:tkgd_length} and \ref{fig:tkgd_pressure}. We note the overall good match between the simulation results and the analytical solution, as well as the convergence of the results towards \ref{length_solution_tkgd} and \ref{pressure_solution_tkgd} when $\ell$ decreases.
Although this problem is relatively simple as the matrix is assumed to be impermeable, it does test the coupling between the global and the local domains and verifies that the phase-field method, even when used only in a vicinity of the crack tip, can still provide accurate predictions of fracture propagation.

\FloatBarrier

We now examine the sensitivity of the results to changes in the subdomain size.  This is accomplished by fixing the sizes of the global and local meshes as well as the regularization length, and varying only $L$ in Figure \ref{fig:2D_schematics_kgd}. In particular, we fix the regularization length to $\ell = 0.13\text{ m}$ and vary the sub-domain size between $15\ell$ and $45\ell$.  

\begin{figure}[!htbp]
\centering
\begin{subfigure}{.5\textwidth}
  \centering
  \includegraphics[width=\linewidth]{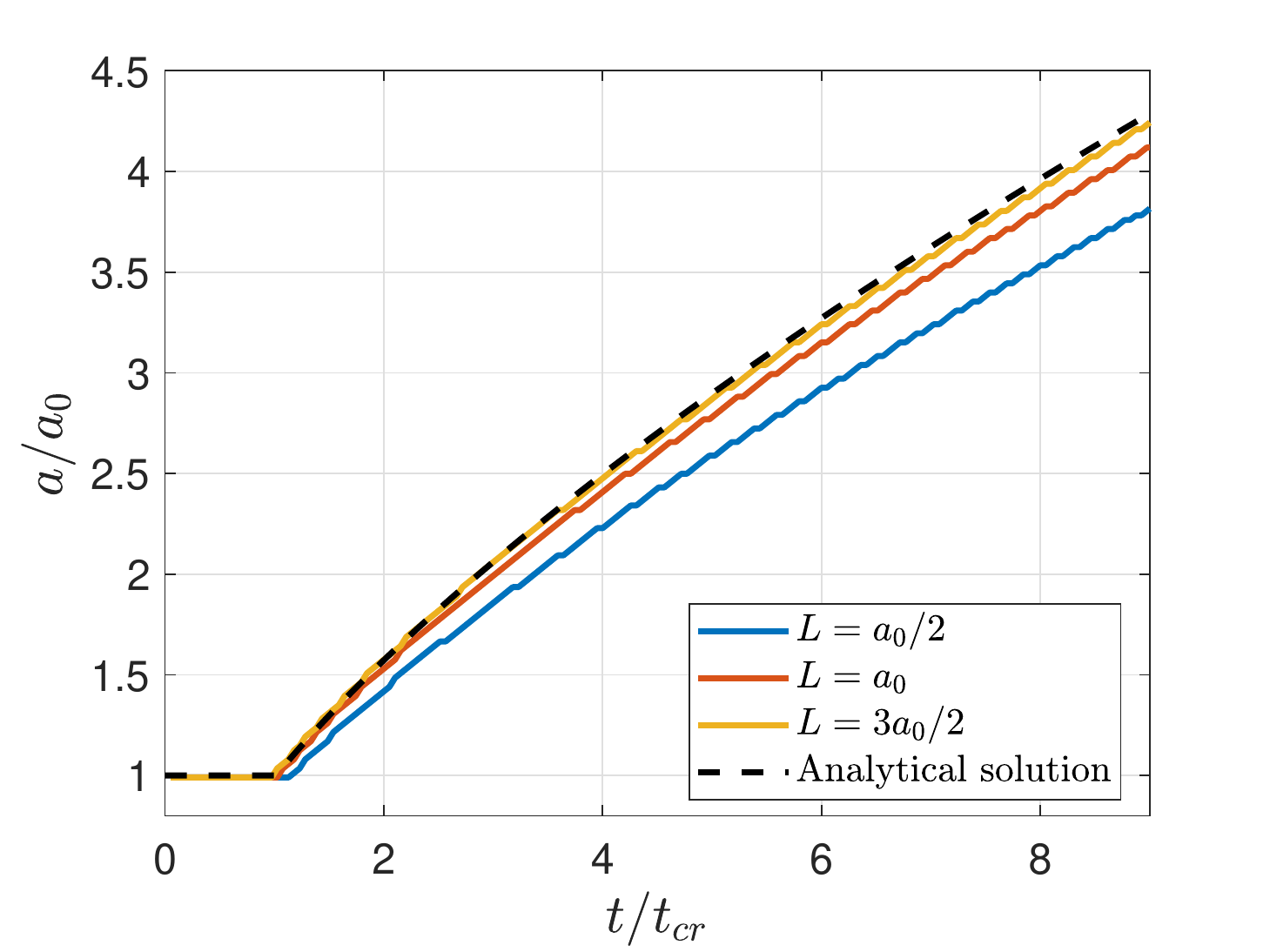}
  \caption{}
  \label{fig:adjusting_L_a}
\end{subfigure}%
\begin{subfigure}{.5\textwidth}
  \centering
  \includegraphics[width=\linewidth]{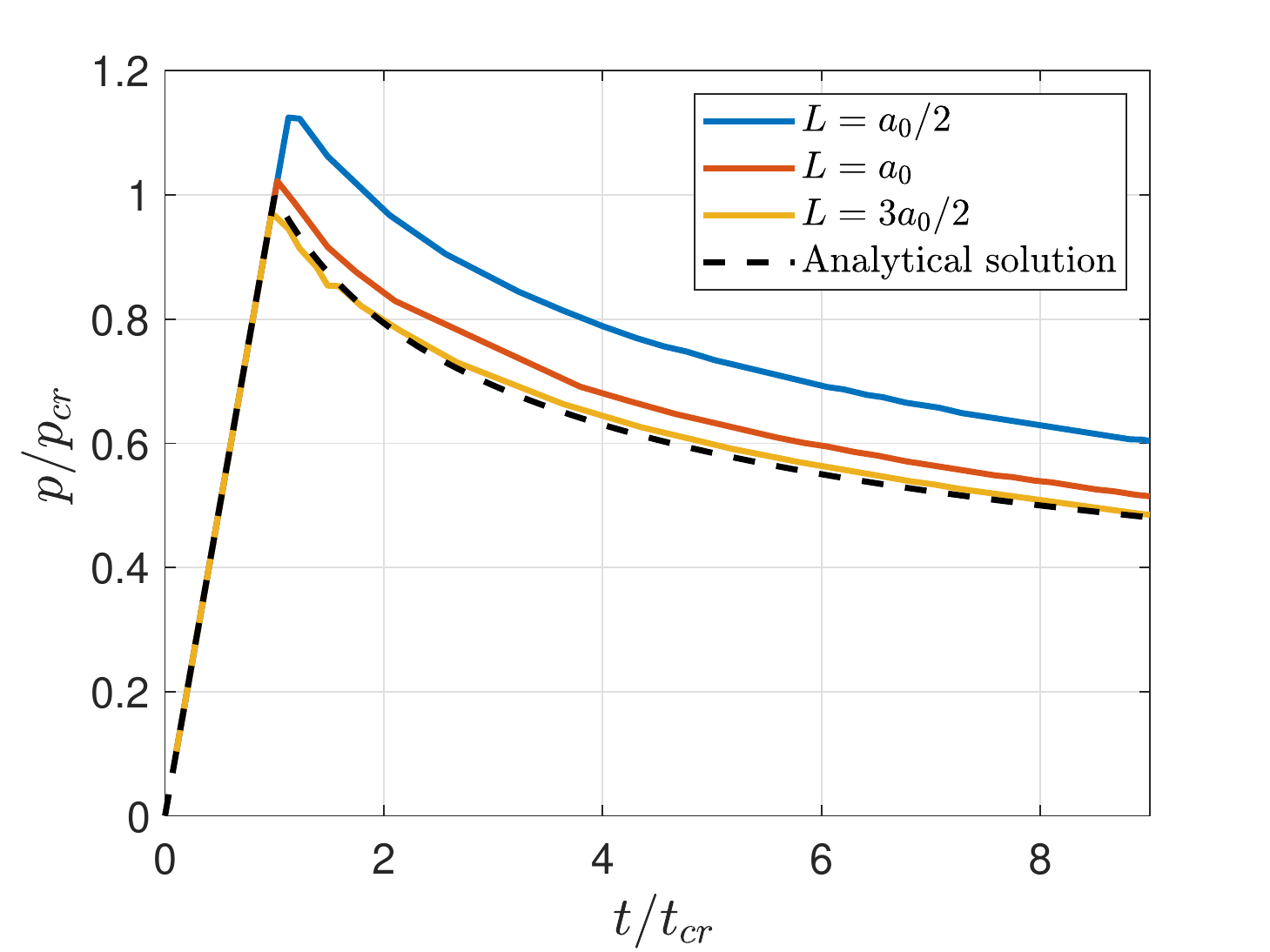}
  \caption{}
  \label{fig:adjusting_L_p}
\end{subfigure}
  \caption{(a) Effects of subdomain size on predicted crack length, for reduced problem. (b) Effects of subdomain size on predicted pressure, for reduced problem. }
  \label{fig:adjusting_L}
\end{figure}

Figures \ref{fig:adjusting_L_a} and \ref{fig:adjusting_L_p} provide the results for the pressure and crack length as a function of time, for the various choices of subdomain size. Table \ref{subdomain_size_table} shows the error in the computation of the crack length relative to the analytical solution.  The error is taken as an average over the time range, starting at the beginning of propagation.

\FloatBarrier

\begin{table}[ht]
\centering
\caption{Effect of subdomain size on the relative error in the crack length. }
\begin{tabular}[t]{lcccc}
\hline
Subdomain size &Relative error   \\
\hline
$L = a_0/2$&11.5\%&\\
$L = a_0$&3.3\%&\\
$L = 3a_0/2$&1.1\%&\\

\hline
\end{tabular}
\label{subdomain_size_table}
\end{table}%

The results indicate that the error reduces as the subdomain size is increased.  This is to be expected, as the subdomain mesh is at a higher resolution than the global mesh.  As a result, there is a trade-off between accuracy and computational cost.  In general, one should select  a subdomain size that strikes a balance between being large enough to capture the crack evolution and small enough to render the overall calculation efficient. 

\subsubsection{Viscosity dominated KGD problem}

We now consider a case in which the choice of parameters gives rise to conditions wherein the energy dissipation is dominated by viscous dissipation.  In particular, we consider the material properties and problem parameters provided in Table~\ref{parameters_vKGD}.  Consistent with \eqref{KGD_group}, these values give rise to $\mathcal{K} = 0.57$, a result that is clearly within the viscosity-dominated regime.   

\begin{table}[ht]
\centering
\caption{Material properties and problem parameters for the viscosity-dominated KGD problem}
\begin{tabular}[t]{lcc}
\hline
&Value &Unit \\
\hline
Young's modulus ($E$)&0.17&GPa\\
Poisson's ratio ($\nu$)&0.20&--\\
Fluid viscosity ($\mu$)&1.0$\times10^{-10}$&$\text{GPa . s}$\\
Energy release rate ($G_c$)&21&$\text{N/m}$\\
Injection rate ($Q$)&1.0$\times10^{-3}$&$\text{m}^2/\text{s}$\\
Initial crack size ($a_0$) &4.0&$\text{m}$\\
\hline
\end{tabular}
\label{parameters_vKGD}
\end{table}%

In this scenario, the toughness of the medium can be neglected, and the fracture evolution is effectively dictated by the motion of the fluid front.  In this case, the pressure varies with time and space over the length of the fracture surface.   Although this problem does not include a pressure field in the matrix,  it bears emphasis that it does require the transfer of the pressure field in the fracture from the global scale to the phase-field problem in the subdomain.  As such, it serves as a test of that aspect of the multi-resolution scheme.  

The presence of dynamic terms in the fluid equation also implies the need for a discretization with sufficiently accurate temporal resolution, and in what follows we examine the sensitivity of the results to the choice of the time step size.  

Once again, consider a computational domain with dimensions $W\times H = 30\text{ m} \times 240\text{ m}$.  The sub-domain size is taken as $L \times L = a_0 \times a_0$, and the regularization length is taken to be $\ell = a_0/15$.  In contrast to what we observed for the toughness regime, we note that the results to the viscosity-dominated problem were found to be largely insensitive to the choice of the regularization length, provided that $\ell < a_0/10$.  This is not surprising, given that the phase-field subproblem only serves to propagate a crack that has effectively zero toughness.

In what follows, we report results using mesh spacings in the local and global domains of $h_{local} = \ell/{4}$ and $h_{global} = 3h_{local}$.  As an initial condition, the pressure field is prescribed to match the analytical solution. We report on the temporal convergence of results obtained using fixed time steps, beginning with $\text{dt}_0 = 0.5$s.  As (discrete) negative pressures arose in our simulations, the phase-field model was modified to account for a tension-compression asymmetry.  In particular, consistent with the model described in Miehe et al. \cite{miehe2010phase}, the strain energy density was split into active and inactive parts, and only the ``tensile" part was degraded with the damage.  

\begin{figure}[!htbp]
\centering
\begin{subfigure}{.5\textwidth}
  \centering
  \includegraphics[width=.99\linewidth]{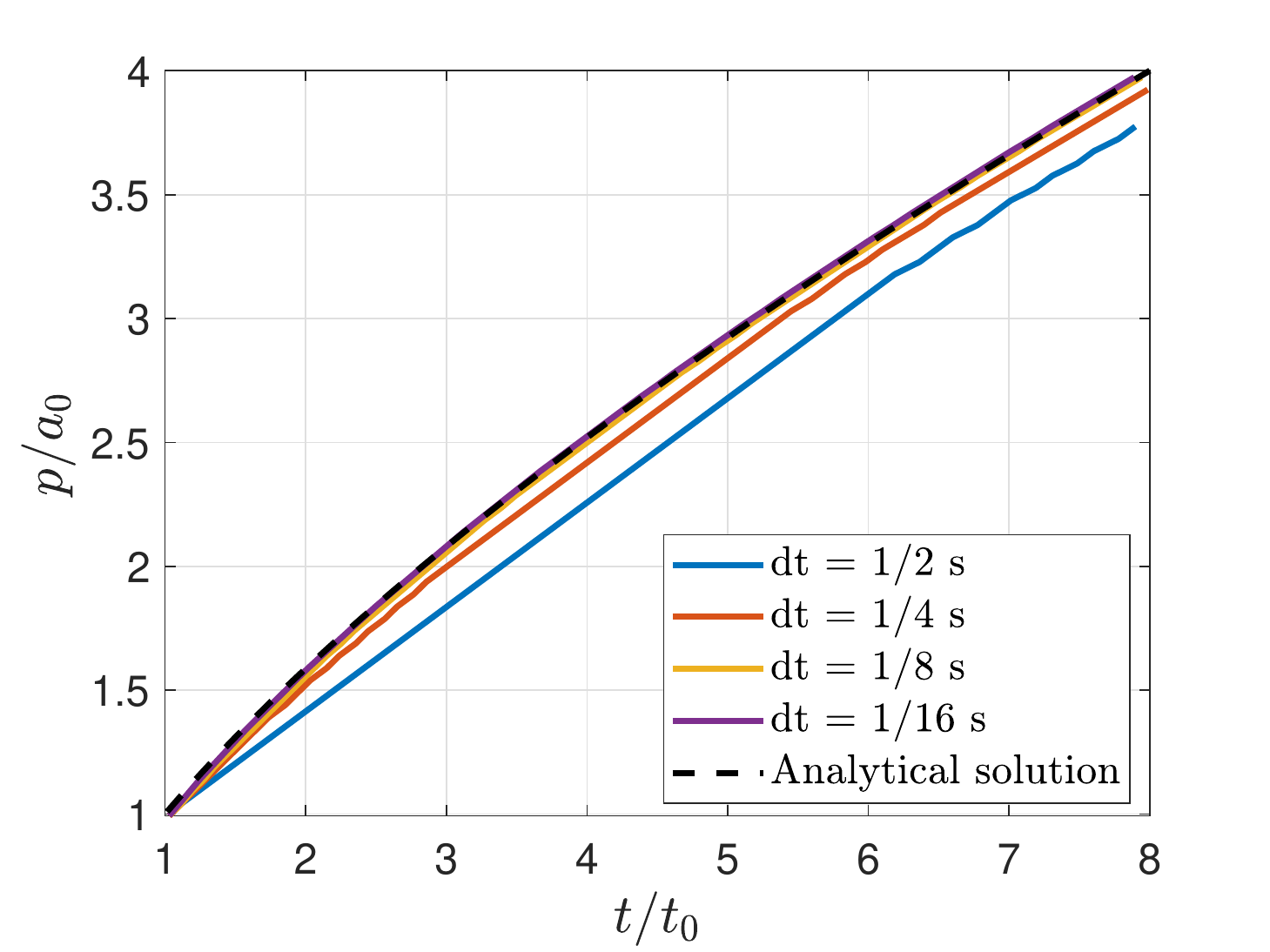}
  \caption{}
  \label{fig:vkgd_length}
\end{subfigure}%
\begin{subfigure}{.5\textwidth}
  \centering
  \includegraphics[width=.99\linewidth]{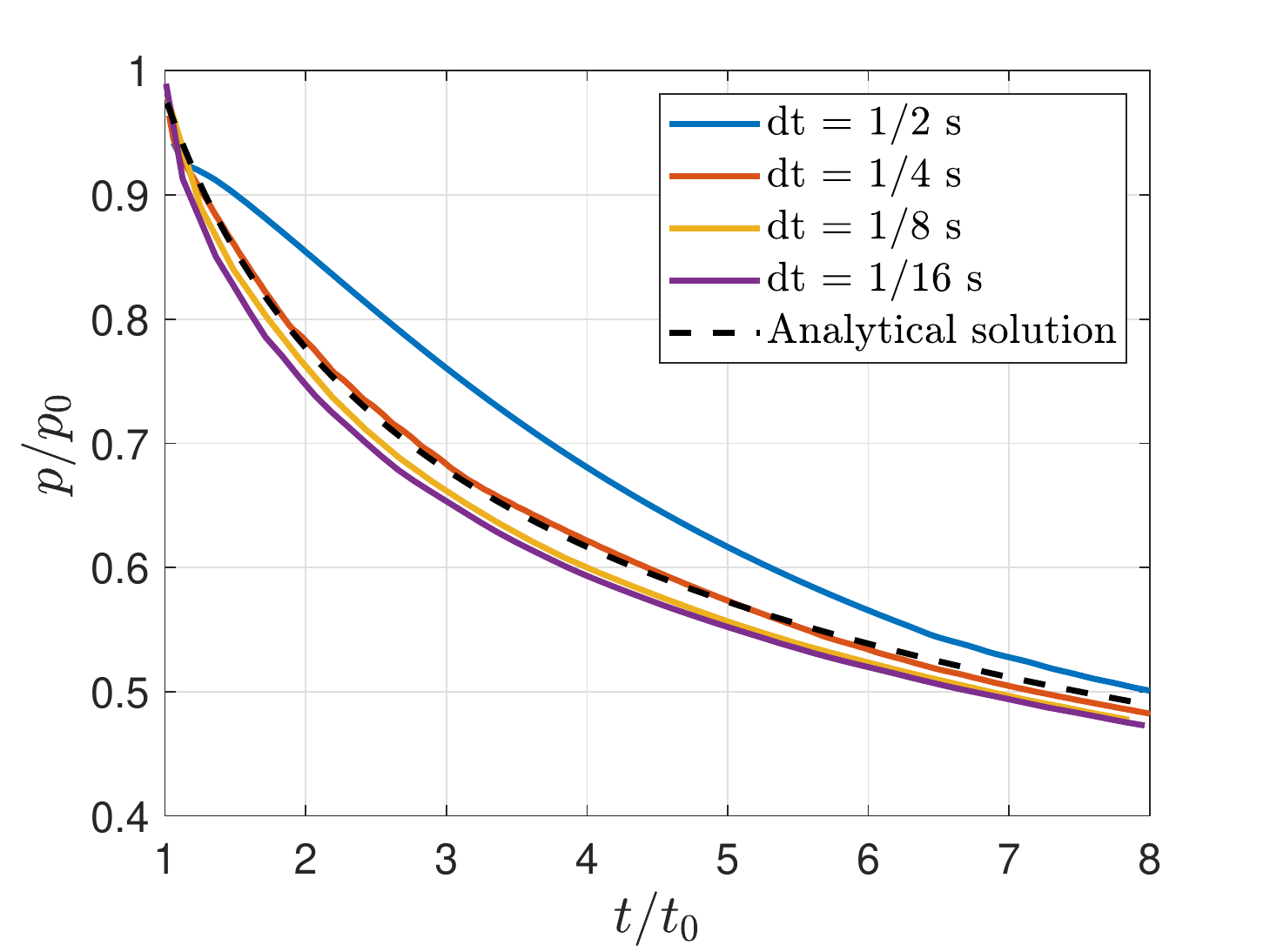}
  \caption{}
  \label{fig:vkgd_pressure}
\end{subfigure}
\caption{Numerical results and analytical solution for the viscosity-dominated KGD problem: (a) crack length vs.\ time; and (b) inlet pressure vs.\ time. }
  \label{fig:vkgd_charts}
\end{figure}

Figure~\ref{fig:vkgd_charts} compares the analytical solutions for the crack length and inlet pressure to the model-based simulation results for a sequence of decreasing time steps.   The initial  time ($t_0 = 8.5 \text{ s}$), initial inlet pressure ($p_0 = 40 \text{ kPa}$) and initial crack size ($a_0 = 4\text{ m}$) were used to render the results dimensionless.   The particular values of $t_0$ and $p_0$ were chosen in order to make our initial state a snapshot of the analytical solution. Overall, the good agreement between the results indicates that the multi-resolution scheme is capable of handling a viscosity-dominated case.    

The results for the crack length show an excellent agreement with the analytical solution as the time step is decreased. The results for the inlet pressure (Figure~\ref{fig:vkgd_charts}b) appear to converge to a trajectory that is slightly offset from the analytical solution as the time step is decreased.  This relatively small discrepancy  can be explained by the presence of a minimum aperture that is assigned to the newly-initialized finite volumes in the flow solver.  As explained in \cite{settgast2017fully}, this is necessary to preserve the stability of the scheme when new crack segments are added. The recent work by Jin et al.~\cite{jin2022robust} proposes a method for removing this parameter by employing partially fractured elements. The incorporation of similar modifications in the context of the multi-resolution scheme is the subject of future work.

\subsection{Poroelastic problem with coupled matrix-fracture flow }

\begin{figure}[!htbp]
    \centering
    \includegraphics[width=.9\linewidth]{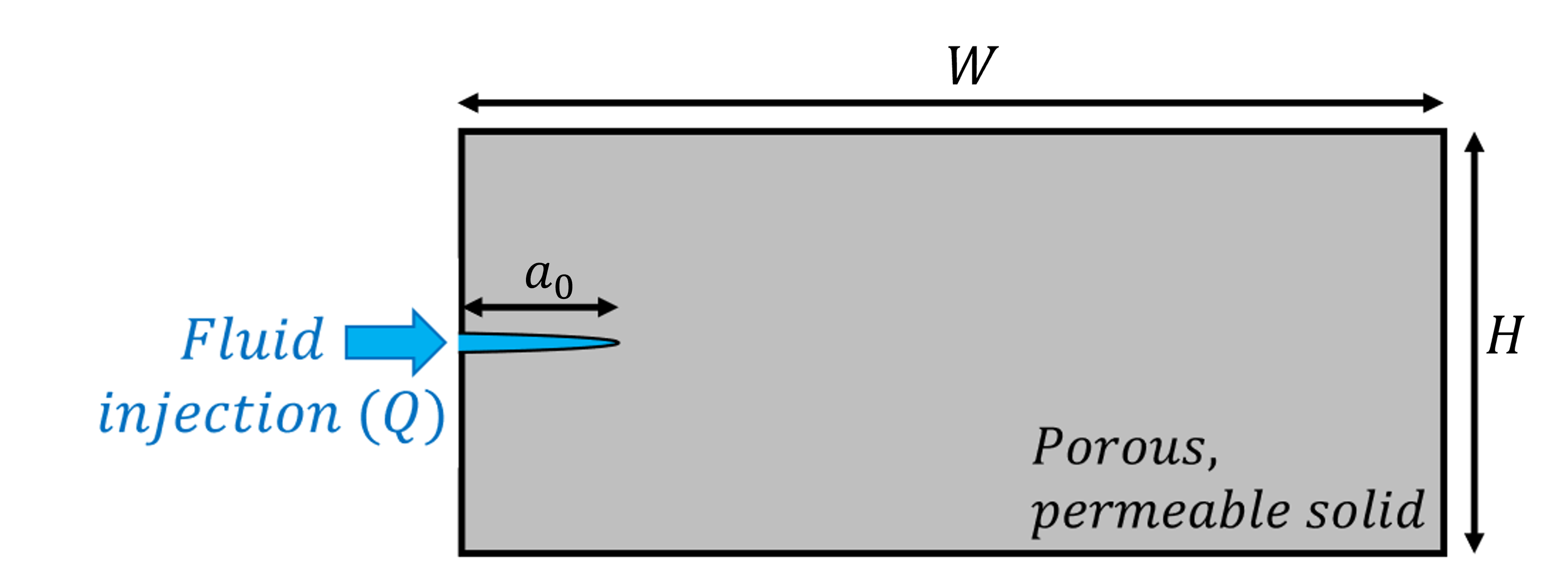}
    \caption{Geometry and notation for the coupled poroelastic-fracture problem.}
    \label{fig:geometry_miehe}
\end{figure}

We now examine a problem in which the matrix is permeable and there is a coupling between the fluid flow in the fracture, fluid flow in the matrix, and crack propagation.  A schematic of the problem is shown in Figure~\ref{fig:geometry_miehe}.   It corresponds to a rectangular domain subject to the injection of a viscous fluid at the mouth of an initial fracture.  The fluid injection rate $Q$ is assumed to be constant, and the displacement in the normal direction is fixed on all sides of the domain.  

Although this problem is relatively simple, to our knowledge an analytical solution is not available.  In what follows, we therefore compare our results to those that are obtained using an existing phase-field method for hydraulic fracture.  In particular, we compare our results to those obtained using the method described by Miehe and Mauthe \cite{miehe2015minimization, miehe2016phase}.  This is arguably one of the simpler methods available for this class of problems, combining the equations for phase-field fracture, Biot's theory of linear poroelasticity in the matrix, and lubrication theory for fluid flow within fractures.

The model assumes that a Darcy model of flow holds in the matrix, with isotropic permeability.  In the fractures, a lubrication flow model is assumed.  The transition between the two flow regimes is effected through the use of a permeability tensor that varies as a function of the crack aperture and orientation.  In particular, the permeability tensor ${\boldsymbol{\kappa}}$ is given by
\begin{equation}\label{miehe_model_basics}
    \boldsymbol{\kappa} = \boldsymbol{\kappa}_0 + d^\xi\dfrac{w^2_n}{12}\left(\textbf{I} - \textbf{n}^d \otimes \textbf{n}^d \right),
\end{equation}

\begin{equation}\label{miehe_model_aperture}
    w_n = h (\textbf{n}^d \cdot \bf{\epsilon} \cdot \textbf{n}^d),
\end{equation}

in which the normalized gradient $\textbf{n}^d = \nabla d/|\nabla d|$ approximates the normal to the crack plane.   In the above, $\boldsymbol{\kappa}_0$ is an isotropic part of the permeability that accounts for the undamaged permeability of the matrix, $w_n$ is the crack's normal aperture, and $\xi$ is a weighting exponent.  The weighting exponent concentrates the effects of the second term in \ref{miehe_model_basics} to areas where $d \approx 1$. Consistent with the work of Miehe and Mauthe \cite{miehe2015minimization, miehe2016phase}, we use $\xi= 50$.

\begin{table}[ht]
\centering
\caption{Material properties for poroelastic problem}
\begin{tabular}[t]{lcc}
\hline
&Value &Unit \\
\hline
Young's modulus ($E$)&16.0&GPa\\
Poisson's ratio ($\nu$)&0.18&--\\
Fluid viscosity ($\mu$)&1.0$\times10^{-12}$&$\text{GPa . s}$\\
Energy release rate ($G_c$)&3.67&$\text{N/m}$\\
Biot coefficient ($\alpha$)&0.79&$\text{ - }$\\
Rock permeability ($\boldsymbol{\kappa_0}$)&1.0$\times 10^{-13}$&$\text{m}^2$\\
\hline
\end{tabular}
\label{material properties miehe}
\end{table}%

\begin{table}[ht]
\centering
\caption{Problem settings}
\begin{tabular}[t]{lcc}
\hline
&Value &Unit \\
\hline
Phase-Field reg. length ($\ell$)&0.2&m\\
Subdomain size ($L$) &4.0&m\\
Domain width ($W$) &30&$\text{m}$\\
Domain height ($H$) &10&$\text{m}$\\
Initial crack size ($a_0$) &4.0&$\text{m}$\\
Injection rate ($Q$)&1.0$\times10^{-3}$&$\text{m}^2/\text{s}$\\
\hline
\end{tabular}
\label{geometry properties miehe}
\end{table}%

\medskip

The material properties used in this problem are given in Table~\ref{material properties miehe}, while the problem and model parameters are provided in Table~\ref{geometry properties miehe}.  The results reported in this section rely on the use of an eigen-decomposition of the strain energy, as described in Miehe et al. \cite{miehe2010phase}.  The subdomain size for the multi-resolution method was chosen to be proportional to the initial crack size, or $a_0 \times a_0$. The problem was discretized spatially using a uniform mesh  in the local domain with  $h_{local} = \ell/4$, while $h_{global} = 3h_{local}$ was used in the global domain. For the temporal discretization, a uniform time step of $\text{dt} = 0.125\text{ s}$ was used.  This level of spatial and temporal discretization was found to yield results that were sufficiently converged.  

An important difference between the phase-field formulation used in Miehe and Mauthe \cite{miehe2015minimization, miehe2016phase} and that employed in the current multi-resolution method concerns the use of pressure-dependent driving forces.  We draw the reader's attention to the terms involving the pressure fields $p_m$ and $p_f$ in the evolution equation \eqref{damage equation} for the damage field.  In our studies of the KGD problem, we found these terms necessary to include in order to obtain sufficiently accurate simulations.  We refer to these terms as ``driving pressures" as the pressure fields contribute directly to the evolution of the damage field.  Importantly, the phase-field formulation of Miehe and Mauthe \cite{miehe2015minimization, miehe2016phase} does not include these terms in the damage evolution equation.   Accordingly, in what follows we find it useful to compare results from Miehe and Mauthe \cite{miehe2015minimization, miehe2016phase} to those obtained using our multi-resolution method with and without the driving pressures.  

\begin{figure}
\centering
\vspace{-\abovedisplayskip}
\begin{subfigure}{.5\textwidth}
  \centering
  \includegraphics[width=\linewidth]{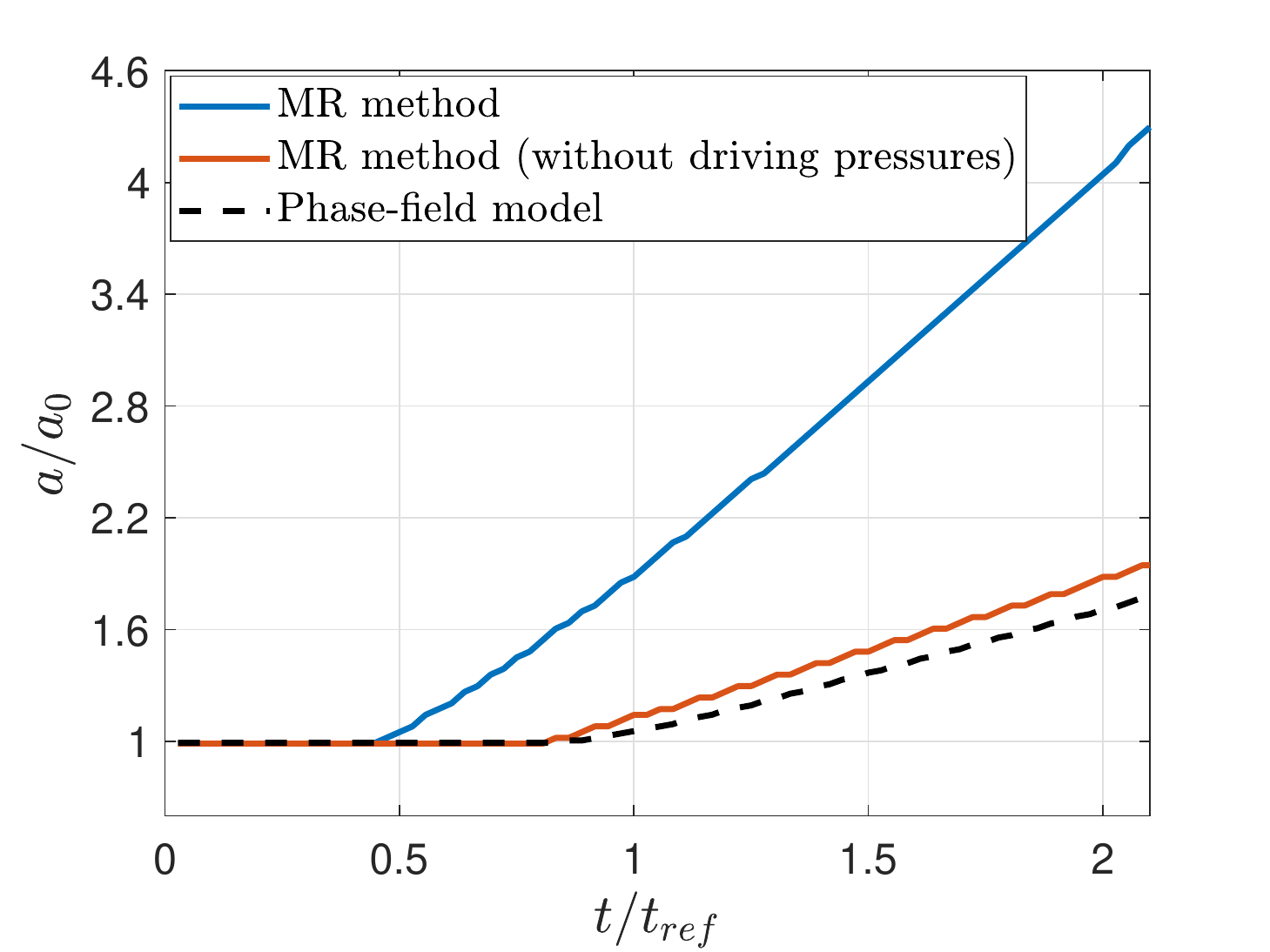}
  \caption{}
  \label{fig:results_size_miehe}
\end{subfigure}%
\begin{subfigure}{.5\textwidth}
  \centering
  \includegraphics[width=\linewidth]{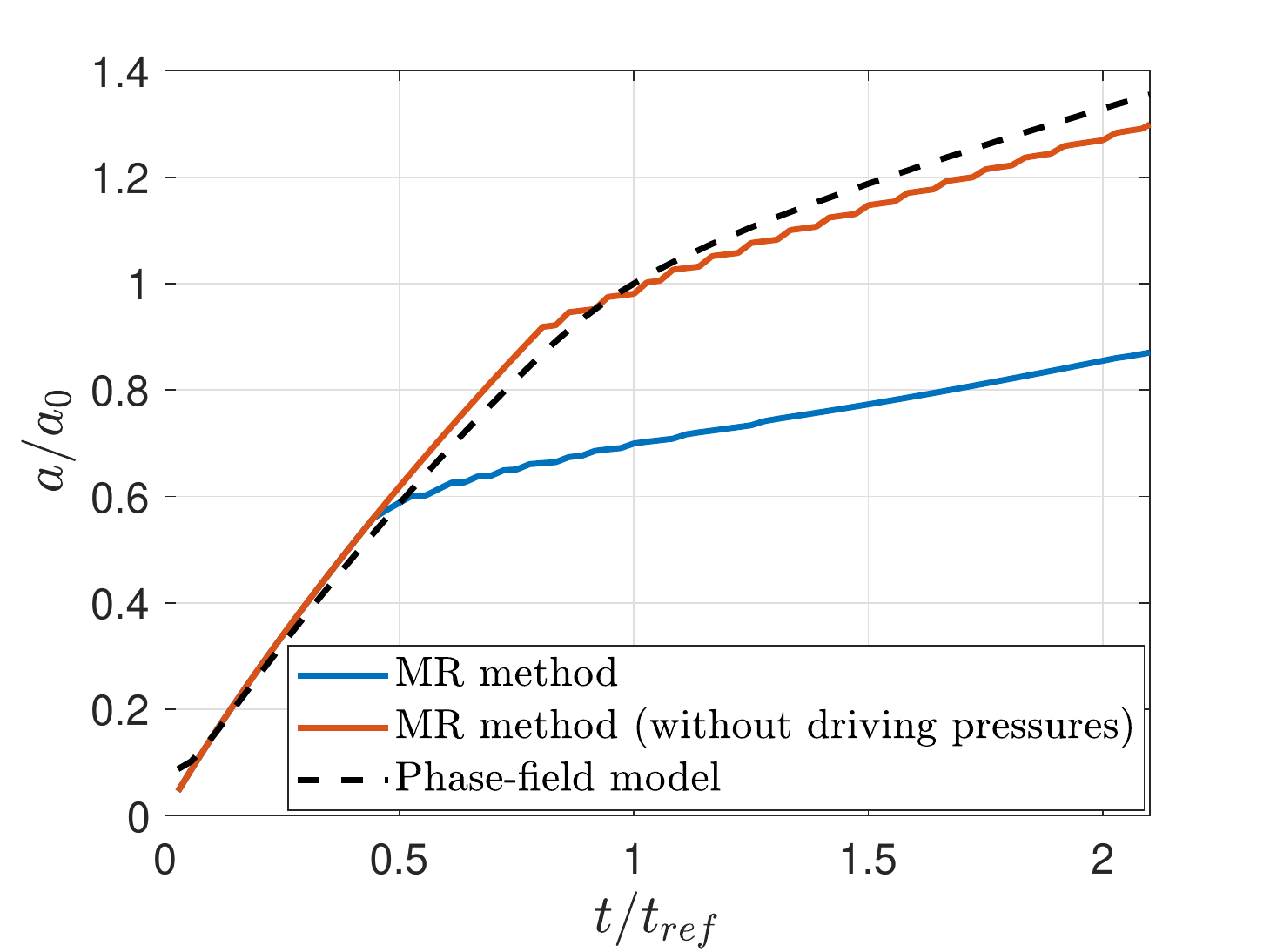}
  \caption{}
  \label{fig:results_pressure_miehe}
\end{subfigure}
  \caption{Results from the fully coupled poroelastic-fracture problem for the multi-resolution method (with and without driving pressures) and a standard phase-field model: (a) crack size over time, (b) inlet pressure.  The time for crack propagation ($t_{ref}$) in the phase-field model and the corresponding inlet pressure ($p_{ref}$) are used as references for the dimensionless charts above.}
  \label{fig:results_charts_miehe}
\end{figure}

\medskip

\begin{figure}
\centering
\vspace{-\abovedisplayskip}
\begin{subfigure}{.5\textwidth}
  \centering
  \includegraphics[width=\linewidth]{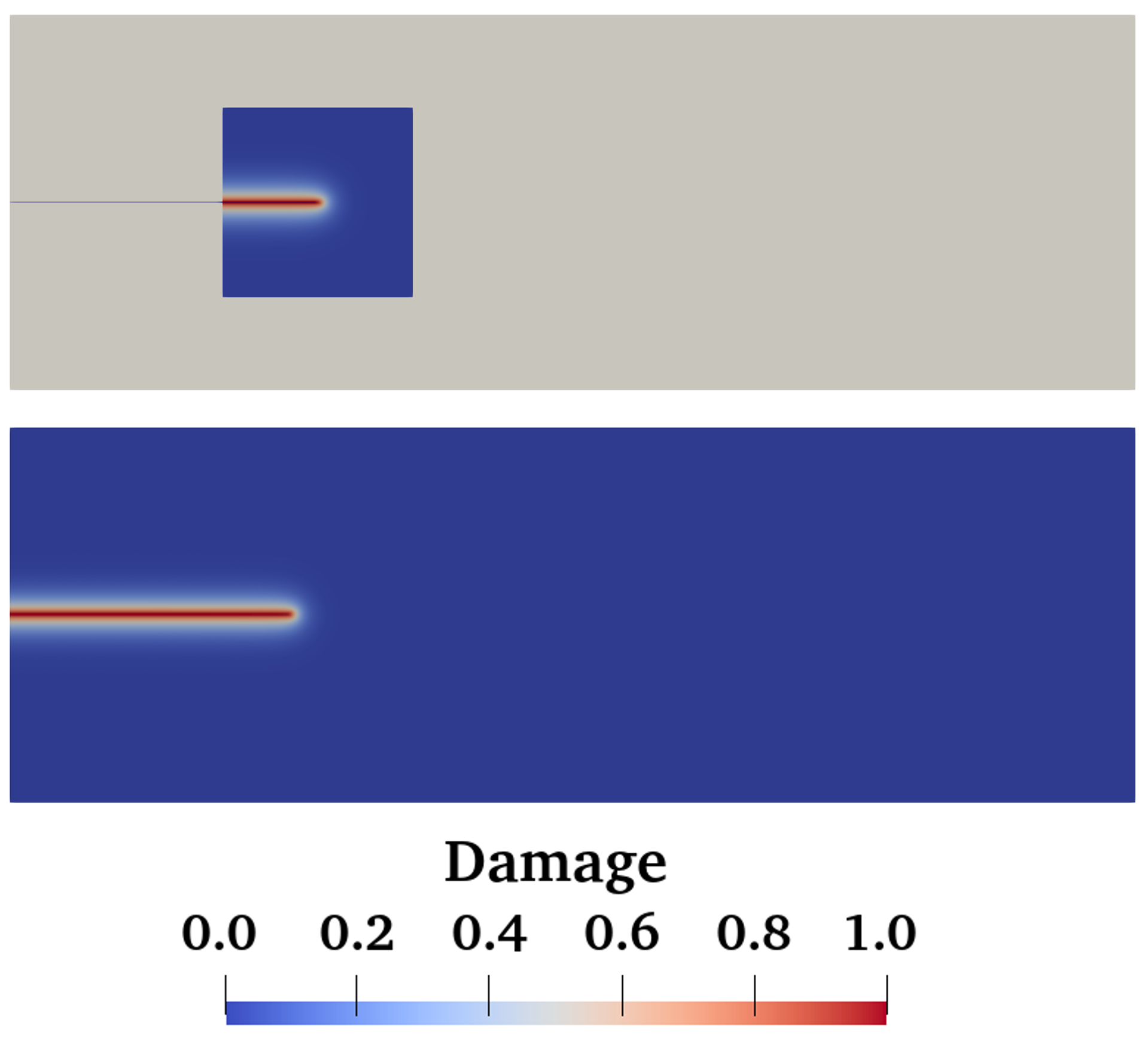}
  \caption{}
  \label{fig:damage_snapshot_p2}
\end{subfigure}%
\begin{subfigure}{.5\textwidth}
  \centering
  \includegraphics[width=\linewidth]{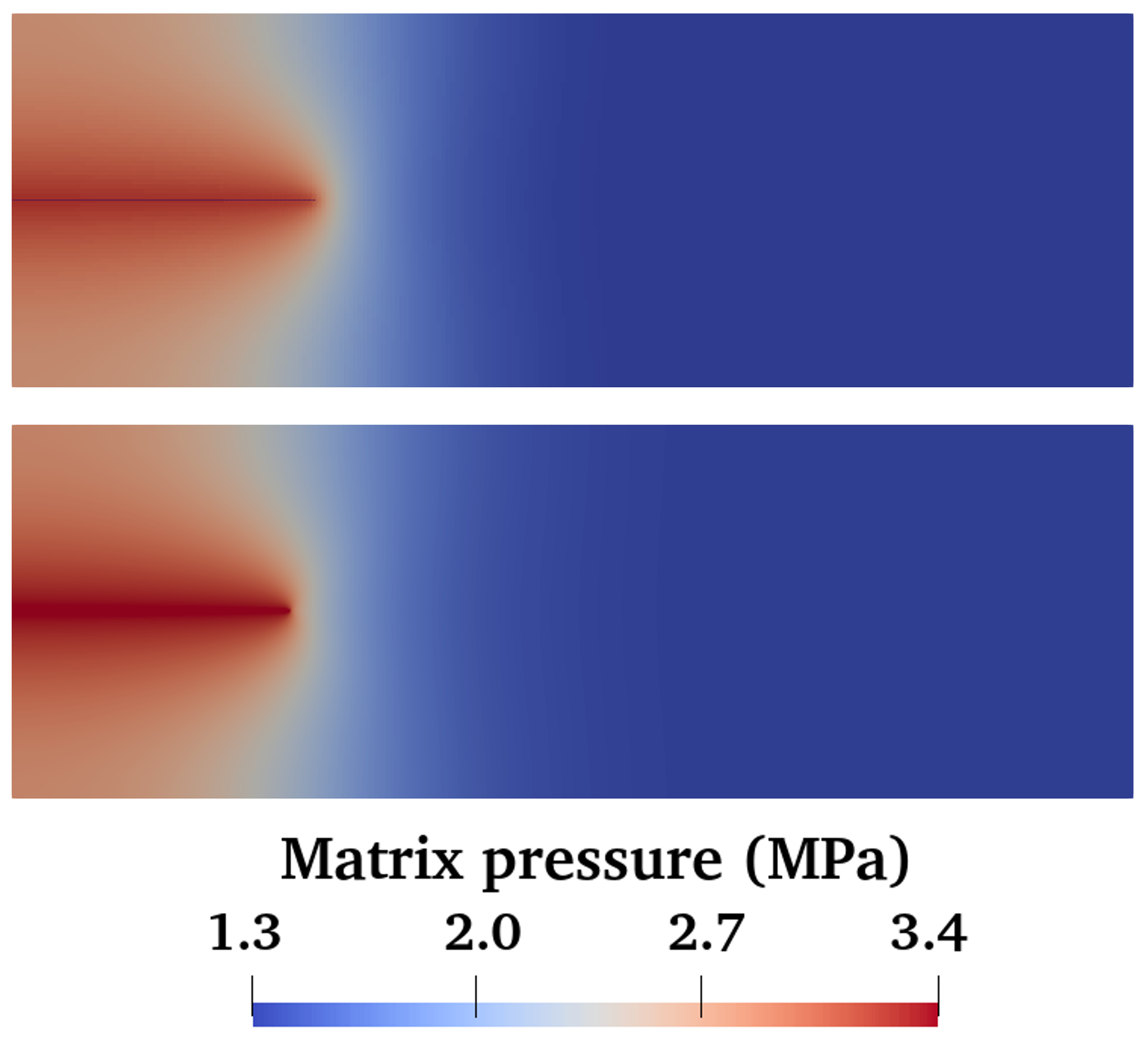}
  \caption{}
  \label{fig:pressure_snapshot_p2}
\end{subfigure}
  \caption{Fields from the coupled poroelastic-fracture problem, taken at the end of the simulation, contrasting results from the multi-resolution scheme without the driving pressures(top row) to those obtained using a standard phase-field model of fracture in poroelastic media (bottom row).  (a) Contour plots of the damage field.   (b) Contour plots of the pressure field.}
  \label{fig:results_fields_miehe}
\end{figure}

Results for the crack length and inlet pressure as a function of time are provided in Figure~\ref{fig:results_charts_miehe}. Contour plots of the damage and pressure fields for the multi-resolution scheme and the full phase-field formulation are provided in Figure~\ref{fig:results_fields_miehe}.
Despite the numerous differences in the models, the results obtained with the multi-resolution compare very favorably to those obtained using our implementation of the phase-field fracture model \cite{miehe2016phase}.  This is particularly the case when the driving pressures are removed from the multi-resolution method, such that the two phase-field models are as close as possible.    At early times the crack remains stationary, as the pressure at the inlet increases.  At some point the pressure near the crack tip reaches a magnitude that is sufficient to give rise to crack propagation.  After crack propagation begins, the rate of pressure increase begins to decrease with time, as crack growth allows for additional fluid to be accommodated within the fracture.    
Clearly the phase-field subproblem with a frozen pressure field that acts effectively as a body force is still capable of simulating crack propagation, even when there is flow in the crack and the matrix. It bears emphasis that the multi-resolution scheme does not need to rely on estimations such as \ref{miehe_model_basics},\ref{miehe_model_aperture} to extract an approximation to the aperture from the regularized crack geometry.  

\subsection{Propagation around an inclusion }

\begin{figure}[!htbp]
    \centering
    \includegraphics[width=.8\linewidth]{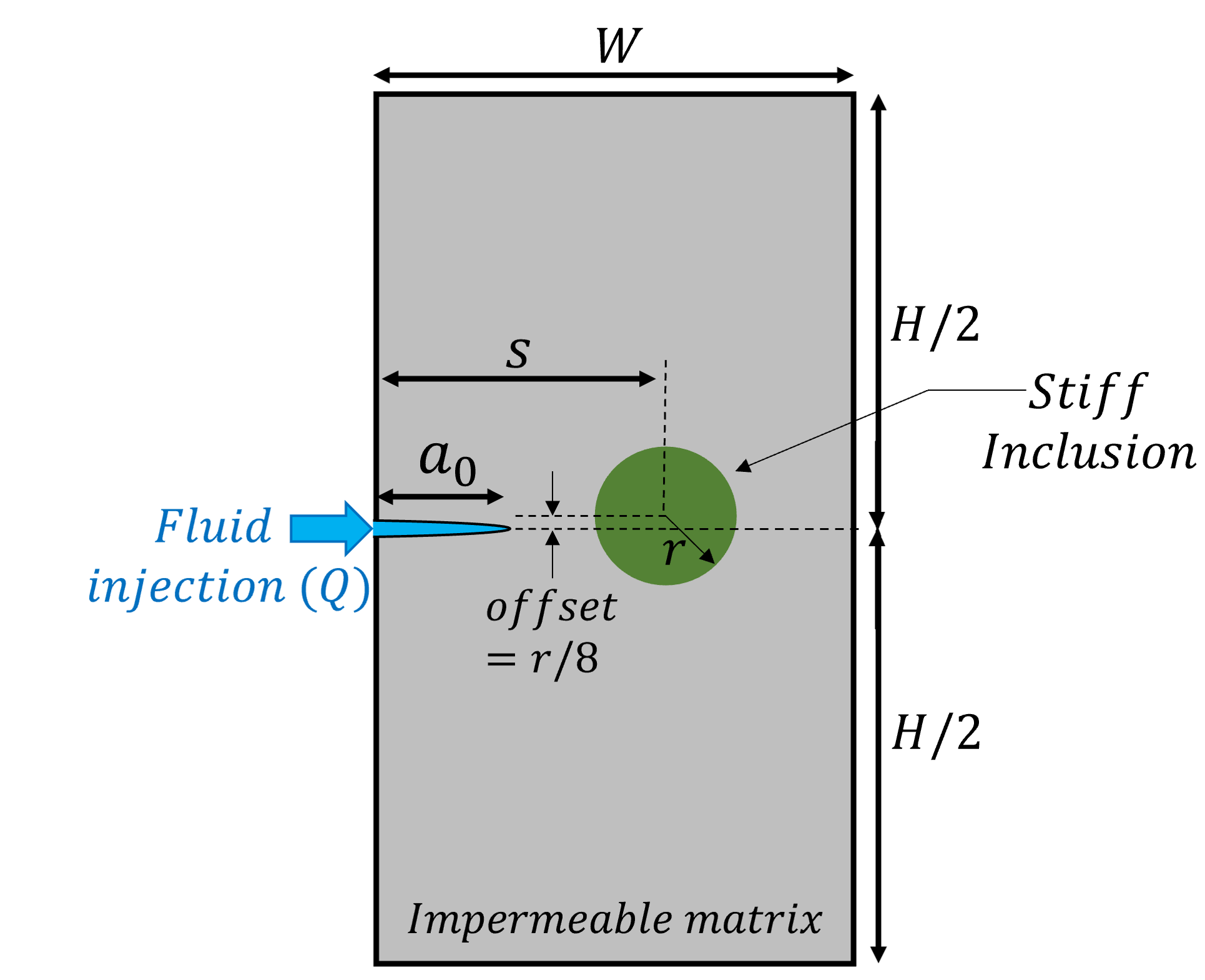}
    \caption{Geometry for the inclusion problem.}
    \label{fig:description_inc_prob}
\end{figure}

We now consider a problem that gives rise to a non-planar crack evolution.   Specifically, we  investigate the effects of a stiff inclusion on the trajectory of a hydraulically-driven fracture in an impermeable medium. The problem setup is shown in Figure \ref{fig:description_inc_prob}. A circular inclusion of radius $r$ is placed in a rectangular domain, at a distance $s$ from the left boundary, and slightly offset from the axis of symmetry. The inclusion is assumed to have properties that are identical to the matrix, with the exception of the Young's modulus.  
An initial crack of size $a_0$ is placed in the left boundary and a fluid is injected at a constant rate $Q$. The complete set of material properties and geometric parameters are listed in Tables \ref{materials_inclusion} and \ref{geometry_inclusion}.

\begin{table}[ht]
\centering
\caption{Material properties for inclusion problem}
\begin{tabular}[t]{lcc}
\hline
&Value &Unit \\
\hline
Young's modulus ($E$)&9.0&GPa\\
Poisson's ratio ($\nu$)&0.25&--\\
Fluid viscosity ($\mu$)&1.0$\times10^{-12}$&$\text{GPa . s}$\\
Energy release rate ($G_c$)&2.5$\times10^{6}$&$\text{N/m}$\\
\hline
\end{tabular}
\label{materials_inclusion}
\end{table}%

\begin{table}[ht]
\centering
\caption{Problem settings}
\begin{tabular}[t]{lcc}
\hline
&Value &Unit \\
\hline
Phase-Field reg. length ($\ell$)&0.4&m\\
Subdomain size ($L$) &4.0&m\\
Domain width ($W$) &20&$\text{m}$\\
Domain height ($H$) &40&$\text{m}$\\
Initial crack size ($a_0$) &4.0&$\text{m}$\\
Injection rate ($Q$)&1.2$\times10^{-1}$&$\text{m}^2/\text{s}$\\
\hline
\end{tabular}
\label{geometry_inclusion}
\end{table}%

Simulations using the multi-resolution scheme were conducted for three different values of inclusion stiffness, with $h_{local} = \ell/6$ in the local domain, and $h_{global} = 3h_{local}$ in the global domain. The objective was to test how the stiffness of the inclusion influenced the crack trajectory. Consider the following two limiting cases.  When the inclusion is just slightly stiffer than the matrix, one would expect the crack trajectory to remain straight. At the other extreme, when the inclusion is much stiffer, one would expect the crack to propagate around it.  

\begin{figure}[!htbp]
\begin{subfigure}{.33\textwidth}
  \centering
  \includegraphics[width=\linewidth]{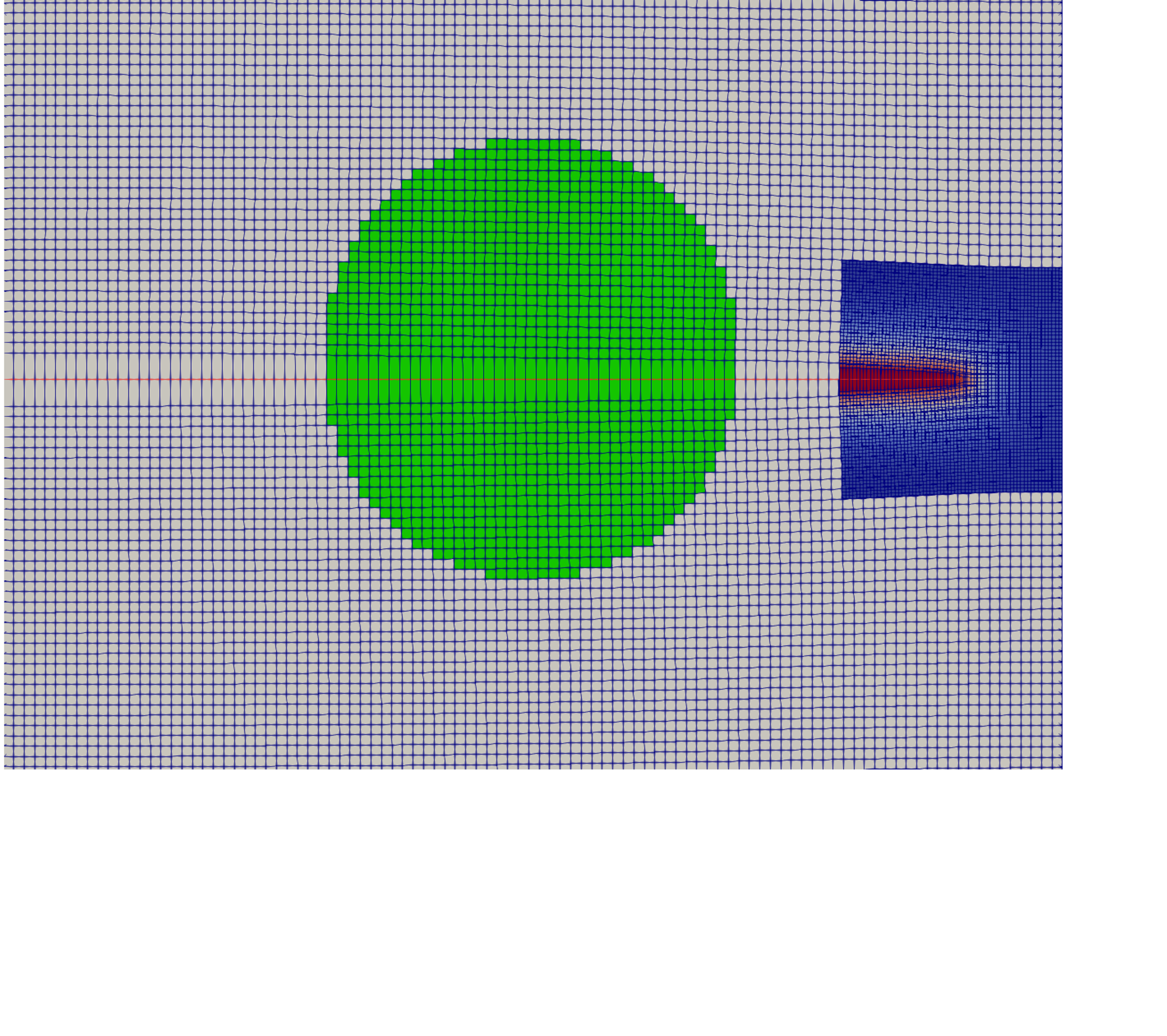}
  \caption{}
  \label{fig:result_2x_inc}
\end{subfigure}%
\begin{subfigure}{.33\textwidth}
  \centering
  \includegraphics[width=\linewidth]{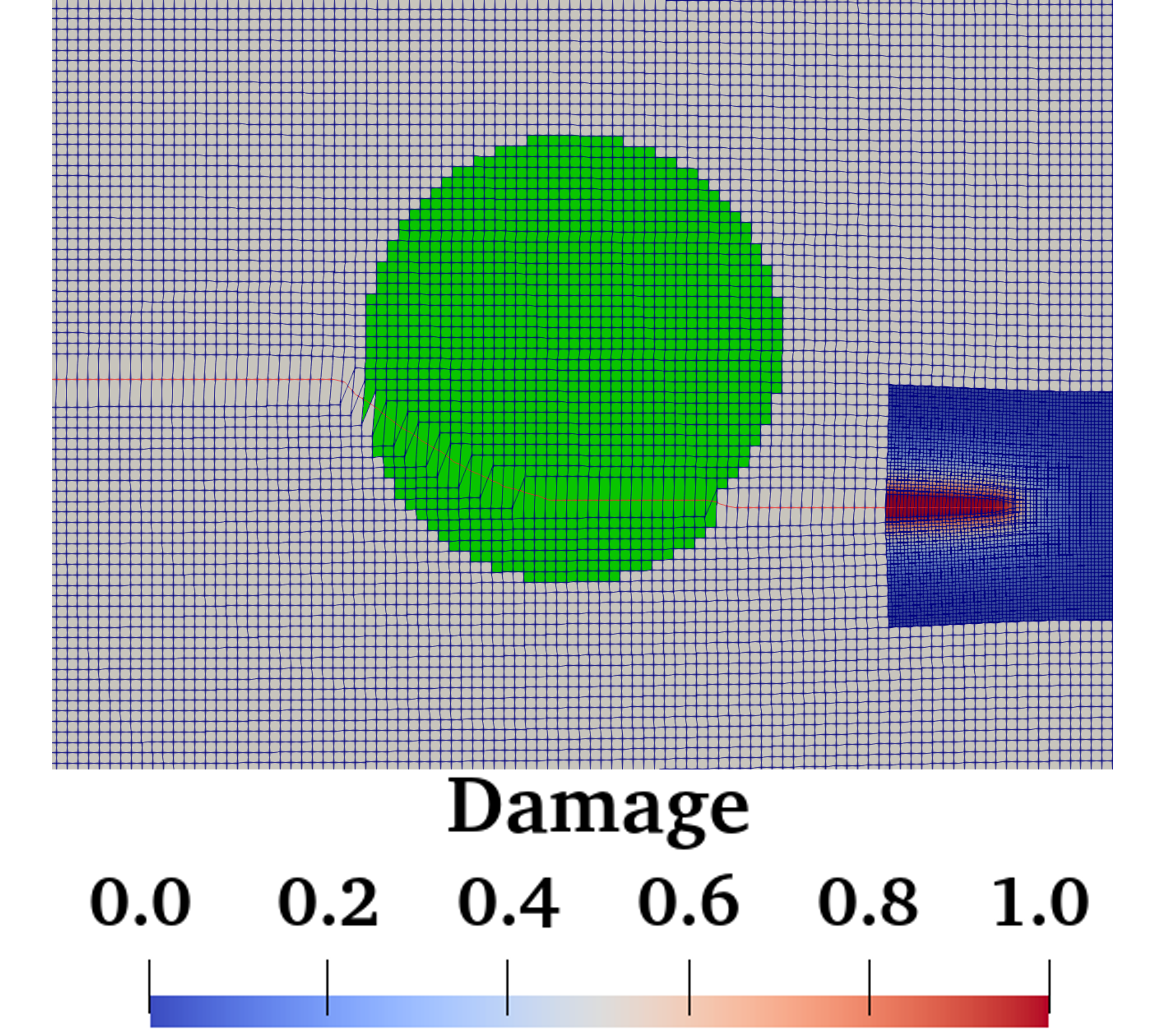}
  \caption{}
  \label{fig:result_7x_inc}
\end{subfigure}%
\begin{subfigure}{.33\textwidth}
  \centering
  \includegraphics[width=\linewidth]{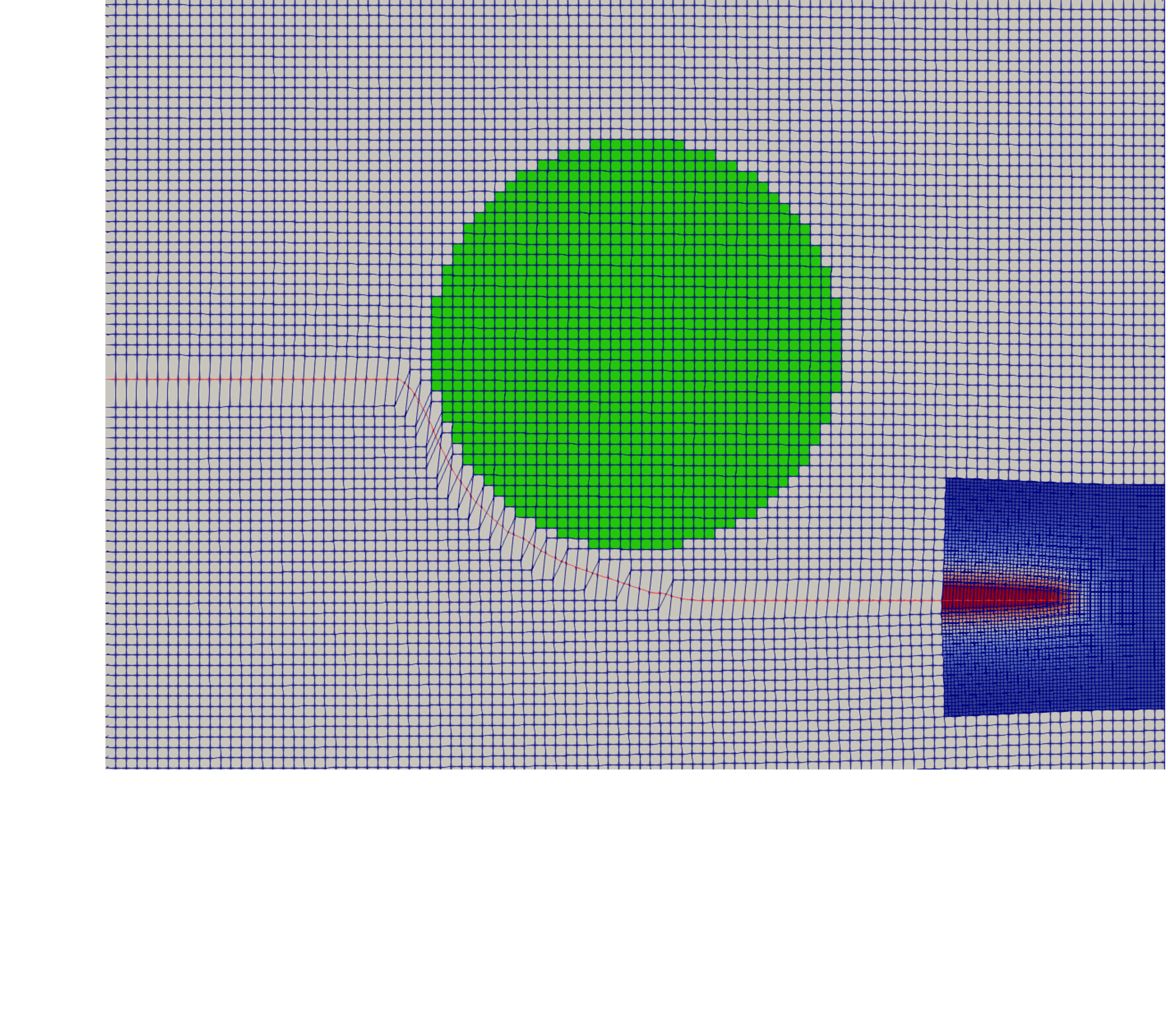}
  \caption{}
  \label{fig:result_15x_inc}
\end{subfigure}
  \caption{Crack paths for the inclusion problem: (a) 2X stiffer inclusion; (b) 7X stiffer inclusion; and (c) 15X stiffer inclusion. } 
  \label{fig:inclusion_paths}
\end{figure}

The crack paths predicted by the multi-resolution scheme are shown in Figure \ref{fig:inclusion_paths}. The deformed meshes are shown with displacements exaggerated to highlight the crack trajectories. As expected, for an inclusion with much higher stiffness, the crack propagates around it. By contrast, when the stiffness of the inclusion is relatively close to that of the matrix (2X case), the trajectory remains nearly straight. Interestingly, for an intermediate value, where the inclusion is 7 times stiffer than the matrix, the crack trajectory gets deflected, but still goes through the inclusion.  

\begin{figure}
\centering
\includegraphics[width=0.5\linewidth]{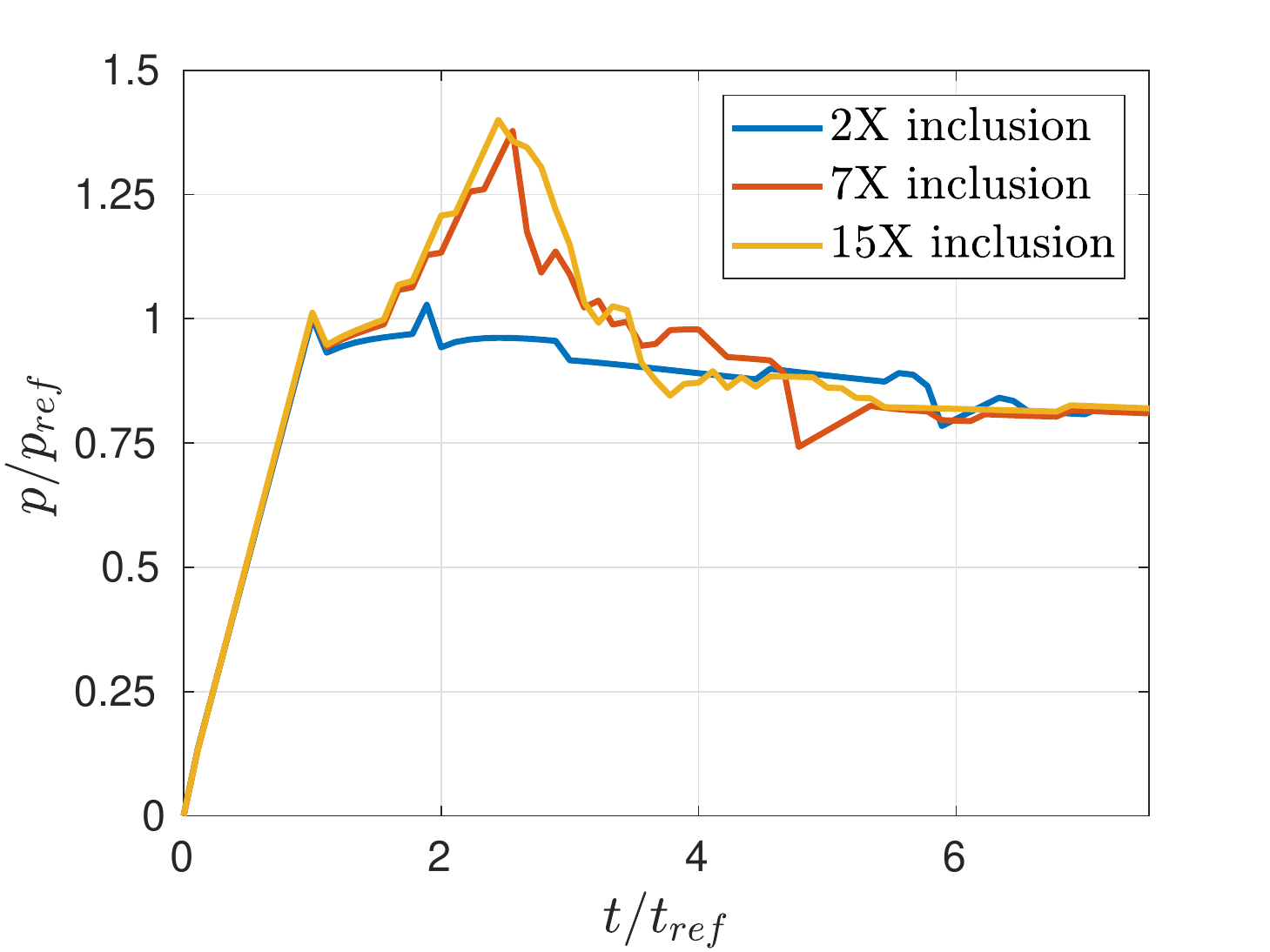}
  \caption{Inlet pressure as a function of time for the three inclusion problems, corresponding to inclusions with different stiffnesses. The reference values of the pressure $p_{ref}$ and time $t_{ref}$ were defined as the minimum pressure that led to fracture growth and the associated time, considering the simulated injection rate.} 
  \label{fig:inclusion_pressures}

\end{figure}

The pressures at the injection point are plotted as a function of time in Figure \ref{fig:inclusion_pressures}. Due to the very high toughness of the material, they are almost uniform over the crack.  For this problem, we did not observe much sensitivity of the results to the choice of time step size.  By analyzing these curves, we can identify several different stages for this problem.  At early times before any crack propagation begins, we observe a linear increase in pressure with time. Then, as we reach a certain pressure, which we denote as $p_{ref} = 60 \text{ MPa}$, straight crack propagation starts and proceeds until the tip approaches the inclusion. Due to the higher stiffness of the inclusion, the crack arrests and pressure builds up again until it reaches a level that is sufficient to allow propagation to continue.  This is accompanied by a pressure drop that is most pronounced for the higher-stiffness inclusion cases.  

This problem highlights several capabilities of the multi-resoltion scheme, such as the simple handling of crack curving as well as changes in the crack speed, including crack arrest. 
Although we are not aware of any analytical solution for problems like this, the results we have obtained appear to make sense qualitatively.  

\FloatBarrier

\section{Summary and conclusions}

This manuscript presents a new approach for developing model-based simulations of hydraulic fracturing. 
The method relies on a global problem that captures flow and deformation fields, and a local problem that captures crack growth with the aid of a phase-field method.  The two problems are coupled through the transfer of displacement and pressure fields, as well as updates to the crack geometry.  This multi-resolution approach allows the phase-field method to be used as a tool to update the crack geometry without the need to explicitly reconstruct the crack aperture from the diffuse representation.  

The accuracy of the multi-resolution approach is evaluated through several numerical examples. These include the well-known KGD problem, for which a good agreement with analytical solutions is obtained in both the toughness and viscosity regimes. A problem with a curved crack trajectory around a stiff inclusion also demonstrates the overall robustness of the approach.  Several areas for future work present themselves. These include an enhancement of the algorithm that tracks the phase-field crack tip, as well as a generalization to three-dimensional problems.

\section*{Acknowledgments}

This work was performed under the auspices of the U.S.\ Department of Energy by Lawrence Livermore National Laboratory under Contract DE-AC52-07NA27344. This research was supported by the Exascale Computing Project (17-SC-20-SC), a collaborative effort of the U.S.\ Department of Energy Office of Science and the National Nuclear Security Administration. The support is gratefully acknowledged. 
\bibliographystyle{elsarticle-num} 
\bibliography{ref_clean.bib}

\end{document}